%% file: acl_latex.tex
\documentclass[11pt]{article}

\usepackage[final]{acl}

\usepackage{array}
\usepackage{amssymb}
\usepackage{times}
\usepackage{pifont}
\usepackage{dsfont}
\usepackage{latexsym}
\usepackage{microtype}
\usepackage[table]{xcolor}
\usepackage{listings}
\usepackage{tcolorbox}
\tcbuselibrary{breakable,listings,skins}
\usepackage{amsmath}
\usepackage{amsthm}
\usepackage{fdsymbol}
\usepackage{bbm}
\usepackage{inconsolata}
\usepackage{float}
\usepackage{placeins}
\usepackage{multirow}
\usepackage{graphicx}
\usepackage{url}
\usepackage{booktabs}
\usepackage{caption}
\usepackage{subcaption}
\usepackage{bbold}
\usepackage{xspace}
\usepackage{tabularx}
\usepackage{makecell}
\usepackage{enumerate}
\usepackage{booktabs}
\usepackage[capitalize,noabbrev]{cleveref}
\usepackage[ruled,vlined]{algorithm2e}
\usepackage[T1]{fontenc}
\usepackage{soul}

\usepackage[utf8]{inputenc}
\DeclareUnicodeCharacter{2192}{$\to$}
\DeclareUnicodeCharacter{2014}{---}
\DeclareUnicodeCharacter{00B3}{\textsuperscript{3}}
\newcommand\eg{\emph{e.g.},\xspace}
\newcommand\ie{\emph{i.e.},\xspace}

\definecolor{redactred}{HTML}{FD3DB5}
\definecolor{actblue}{HTML}{3C78D8}
\newcommand{\papertitle}{\textbf{\textcolor{redactred}{\textsc{Red}}\textcolor{actblue}{\textsc{Act}}}\xspace}
\DeclareRobustCommand{\redacttitleicon}{\raisebox{-0.30em}{\includegraphics[height=1.3em]{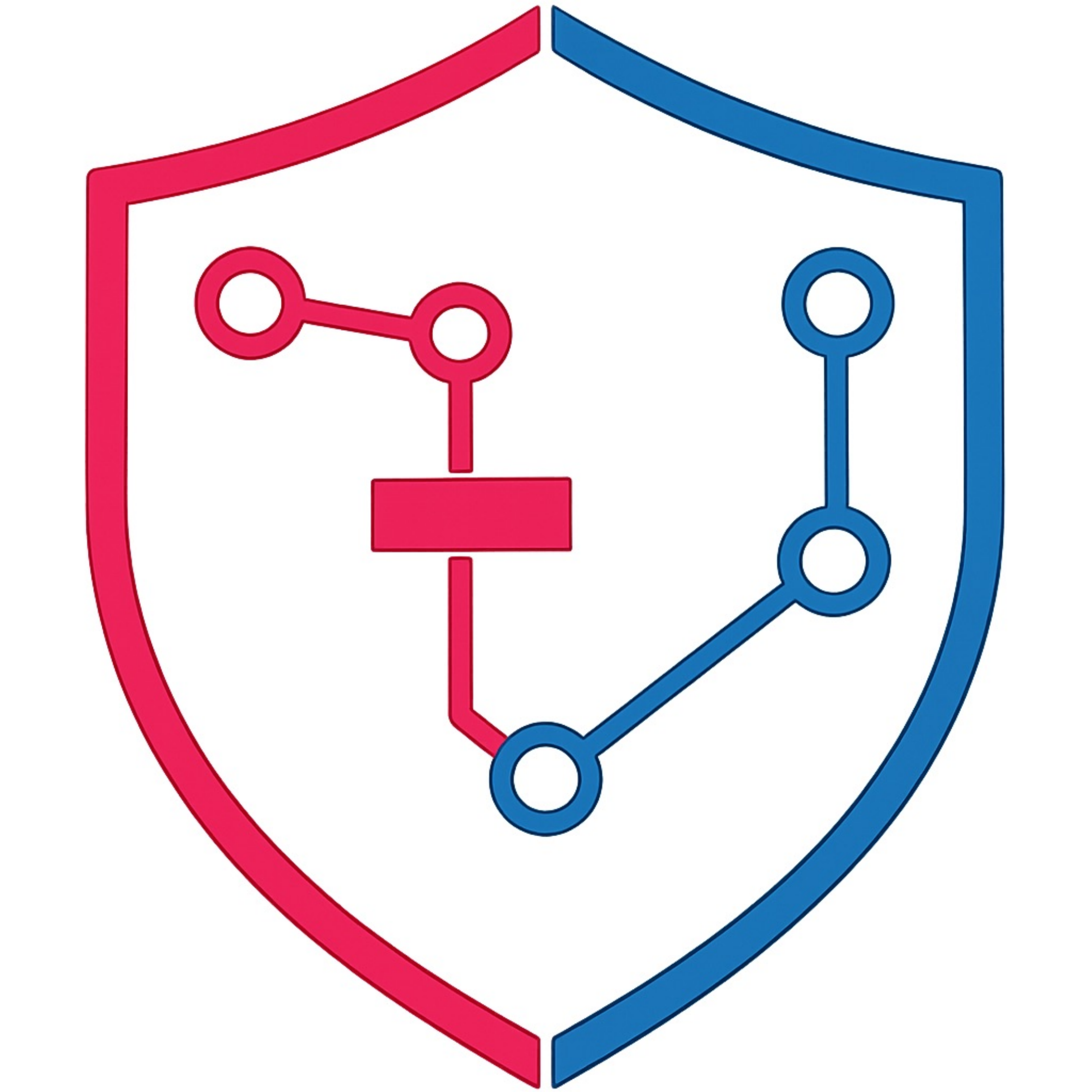}}\hspace{0.16em}}
\newcommand{\benchname}{\textsc{CapTraceBench}\xspace}

\definecolor{srblue}{HTML}{EEE6F9}
\definecolor{ssrgreen}{HTML}{E8F5EB}
\definecolor{deltaGreen}{HTML}{2E7D32}
\definecolor{ppkpurple}{HTML}{EEE6F9}
\definecolor{diffeasy}{HTML}{EEF7F8}
\definecolor{diffmedium}{HTML}{DDEFF4}
\definecolor{diffhard}{HTML}{C9E1EE}
\definecolor{diffavg}{HTML}{E6EAF0}

\definecolor{deltaup}{HTML}{C62828}
\definecolor{deltadown}{HTML}{2E7D32}
\definecolor{deltazero}{HTML}{667085}

\definecolor{rewriteorange}{HTML}{E67E22}
\definecolor{watermarkpurple}{HTML}{7E57C2}
\DeclareRobustCommand{\rewriteicon}{\raisebox{-0.12em}{\includegraphics[height=1.05em]{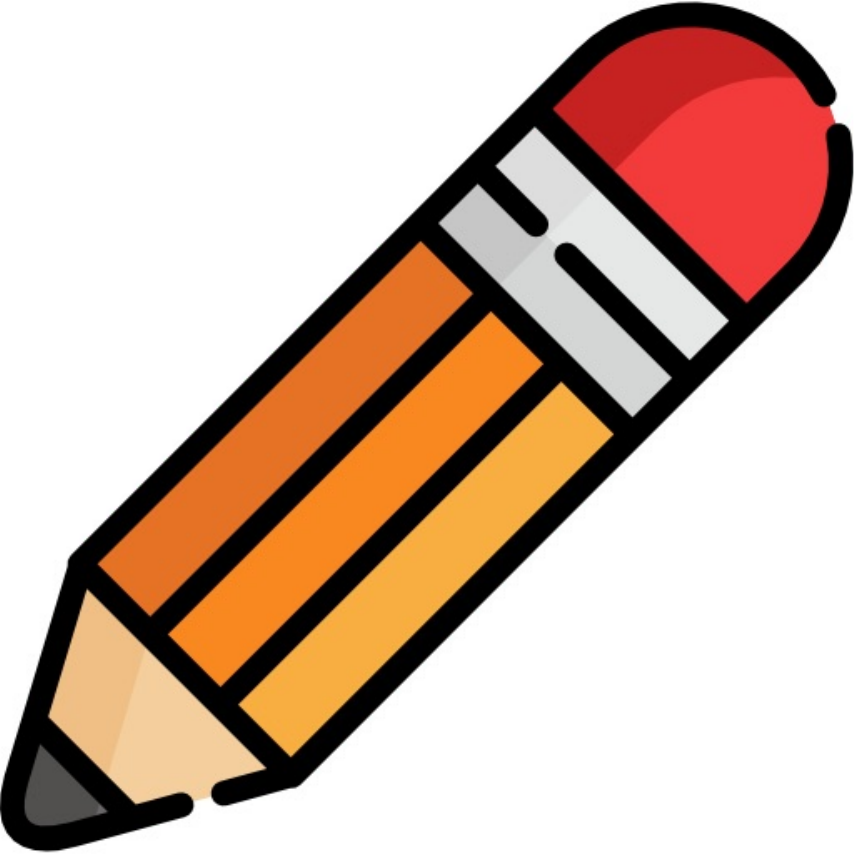}}}
\DeclareRobustCommand{\wmkicon}{\raisebox{-0.12em}{\includegraphics[height=1.05em]{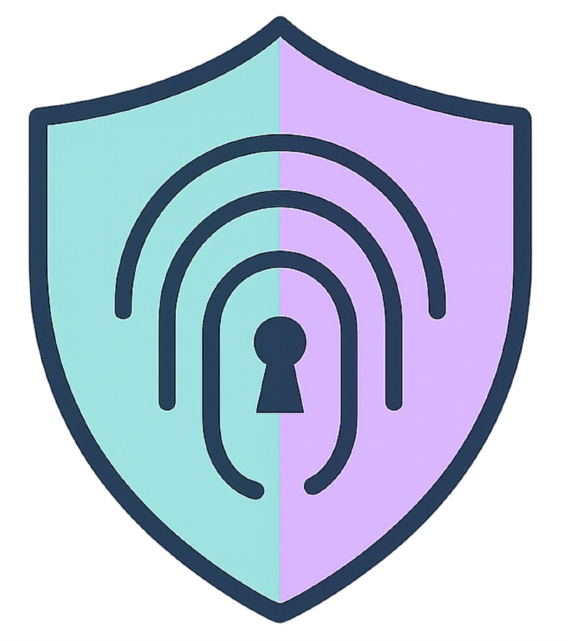}}}

\ifdefined\metriccolwidth\else\newlength{\metriccolwidth}\fi
\ifdefined\settingcolwidth\else\newlength{\settingcolwidth}\fi
\ifdefined\diffcolwidth\else\newlength{\diffcolwidth}\fi
\ifdefined\diffboxwidth\else\newlength{\diffboxwidth}\fi
\newcolumntype{S}{>{\centering\arraybackslash}m{\settingcolwidth}}
\newcolumntype{D}{>{\centering\arraybackslash}m{\diffcolwidth}}
\newcolumntype{M}{>{\centering\arraybackslash}m{\metriccolwidth}}

\newcommand{\capbox}[2]{%
  \begingroup
  \setlength{\fboxsep}{0pt}%
  \colorbox{#1}{%
    \strut\hspace{2.1pt}#2\hspace{2.1pt}%
  }%
  \endgroup
}

\newcommand{\TableColorBox}[3]{%
  \begingroup
  \setlength{\fboxsep}{0pt}%
  \colorbox{#1}{\makebox[#2][c]{\strut #3}}%
  \endgroup
}

\newcommand{\MetricHeadBox}[2]{%
  \hspace*{-\tabcolsep}%
  \TableColorBox{#1}{\dimexpr\metriccolwidth+2\tabcolsep\relax}{#2}%
  \hspace*{-\tabcolsep}%
}

\newcommand{\SRHead}{\MetricHeadBox{srblue}{SR}}
\newcommand{\SSRHead}{\MetricHeadBox{ssrgreen}{SSR}}

\newcommand{\DiffCell}[2]{%
  \TableColorBox{#1}{\diffboxwidth}{#2}%
}
\newcommand{\DiffEasy}{\DiffCell{diffeasy}{Easy}}
\newcommand{\DiffMedium}{\DiffCell{diffmedium}{Medium}}
\newcommand{\DiffHard}{\DiffCell{diffhard}{Hard}}
\newcommand{\DiffAvg}{\DiffCell{diffavg}{Avg.}}

\SetKwInOut{Input}{Input}
\SetKwInOut{Output}{Output}
\SetKwComment{Comment}{$\triangleright$\ }{}

\title{\redacttitleicon\papertitle: \textcolor{redactred}{Red}acting \textcolor{actblue}{A}gent \textcolor{actblue}{C}apability \textcolor{actblue}{T}races for Procedural Skill Protection}

\author{
\textbf{Shuwen Xu$^{1,2}$ ~~Zhitao He$^{1}$ ~~Yi R. (May) Fung$^{1}$} \\
$^{1}$Hong Kong University of Science and Technology \\
$^{2}$University of Chinese Academy of Sciences \\
\texttt{sxucn@connect.ust.hk}
~~~\texttt{yrfung@ust.hk}
}

\begin{document}
\maketitle

\input{sections/0_abstract}
\input{sections/1_introduction}
\input{sections/2_problem_setting}
\input{sections/3_benchmark}
\input{sections/4_method}
\input{sections/5_experiments}
\input{sections/6_related_work}
\input{sections/7_conclusion}
\newpage
\input{sections/limitation_ethics}

\bibliography{custom}
\newpage
\input{sections/appendix}

\end{document}

%% file: sections/0_abstract.tex
\begin{abstract}
Users rely on execution traces to observe agent behavior, diagnose failures, and ensure accountability.
These traces contain rich procedural detail, including tool invocations, intermediate decisions, and error-recovery logic.
Yet this detail can expose private procedural skills, allowing downstream methods to recover key formulas, thresholds, and strategies without access to model weights or skill files.
To quantify this risk and evaluate protection, we construct \benchname, a benchmark of 75 specialized long-horizon tasks and 154 curated skills across seven domains.
We also introduce \papertitle \footnote{\url{https://github.com/XuShuwenn/RedAct}}, a protected trace release framework that localizes protected key information, rewrites traces while preserving verifier-critical evidence, and embeds behavioral watermarks for downstream provenance analysis.
Across representative trace reuse methods, \textsc{RedAct} reduces normalized skill transfer (NST) from 44.7--67.1\% on raw traces to below the no-skill baseline, while preserving audit evidence. Its standalone behavioral watermarks reach 93.6--100.0\% true detection with a false alarm rate of at most 1.9\%.
These results frame public agent traces as security interfaces and show that selective redaction can reduce procedural capability leakage without removing audit evidence.
\end{abstract}

%% file: sections/1_introduction.tex
\section{Introduction}

\begin{figure}[t]
\centering
\includegraphics[width=\columnwidth]{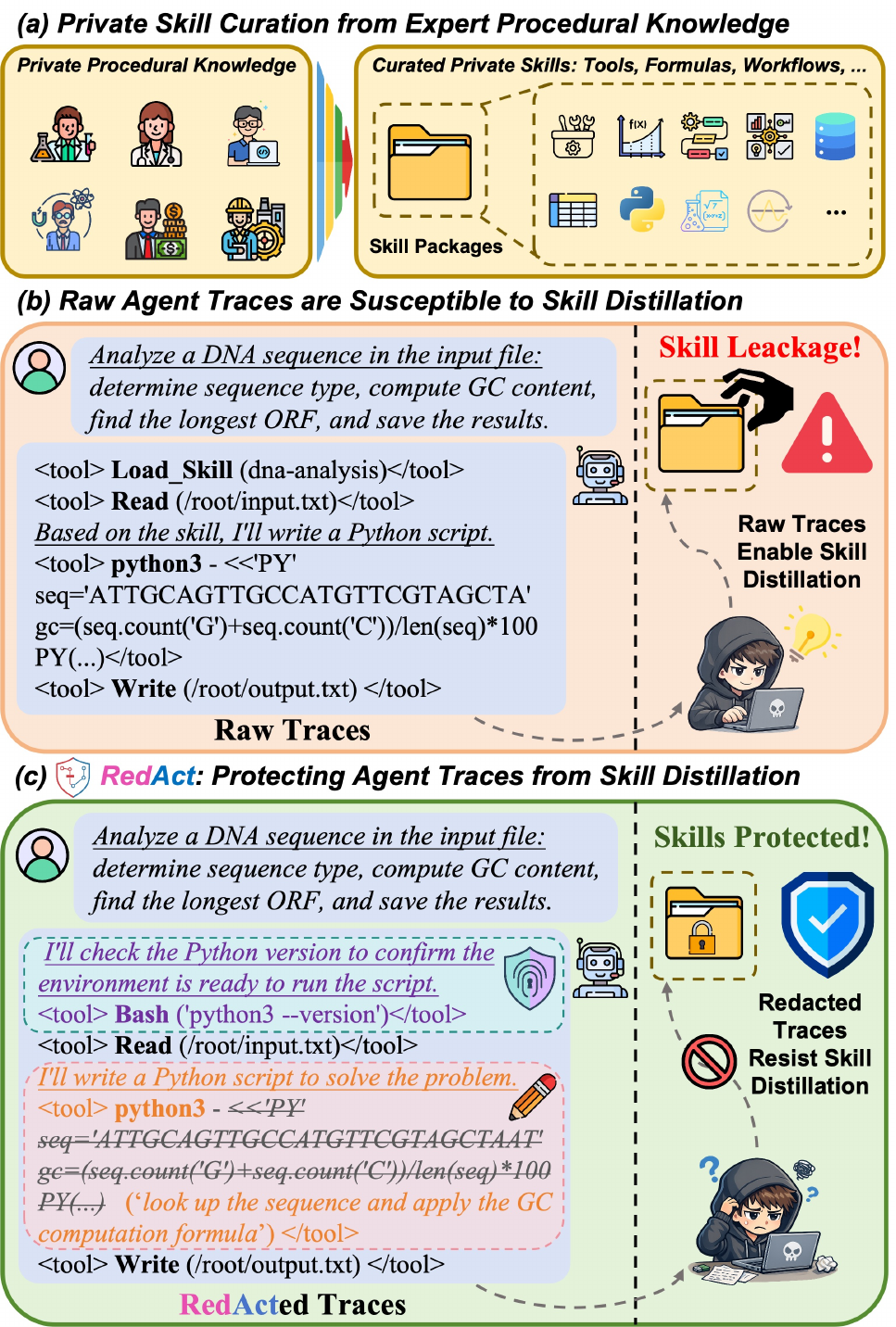}
\vspace{-0.24in}

\caption{
Problem motivation of our work.
Raw agent traces may reveal \textit{reusable private skills} and enable \textit{skill distillation}. 
\papertitle protects released traces via \textcolor{rewriteorange}{selective rewriting}~(\rewriteicon) and \textcolor{watermarkpurple}{behavioral watermarking}~(\wmkicon).
}
\label{fig:overview}
\vspace{-0.26in}
\end{figure}

As tool-using agents transition from research prototypes to deployed systems~\citep{yao2023react, schick2023toolformer, shinn2023reflexion, wang2023voyager}, expert teams in biomedicine, finance, and engineering increasingly build agents around \textit{proprietary data, domain models, and expert-refined workflows}. 
These systems often rely on \textit{\textbf{agent skills}}: structured packages of procedural knowledge encompassing reusable workflows, code templates, tool-use routines, validation scripts, and domain heuristics~\citep{li2026skillsbench, wang2026skillx}. 
A drug-screening pipeline or a financial valuation workflow, for instance, may encode proprietary formulas, calibrated thresholds, multi-stage validation protocols, and recovery strategies.
Execution traces are often released to support transparency, debugging, and user trust; however, the same traces can expose the protected procedures that underlie agent effectiveness~\citep{ni2026trace2skill, chen2026skillcraft}.
This creates a fundamental tension between \textit{trace auditability} and \textit{procedural skill protection}.

Recent work on \textit{\textbf{skill learning and evolution}} has demonstrated that \textit{agent experience can be distilled into reusable skills}, which substantially improves downstream task performance across diverse domains~\citep{wang2026skillx, jiang2026xskillcontinuallearningexperience, ni2026trace2skill, li2026skillsbench, wang2023voyager}.
This transforms trace release into a \textit{procedural asset-protection problem}: a downstream party with access only to released traces may recover proprietary formulas, threshold configurations, tool-call sequences, and validation routines.
We therefore ask: \textbf{\textit{how can useful agent traces be released without exposing reusable protected skills, while still preserving evidence of downstream reuse?}}

We formalize this problem as \textit{black-box trace disclosure}: downstream reuse methods observe only released artifacts, without access to private skill files, model weights, or hidden states.
The central risk is not answer copying, but the recovery of reusable protected procedures applicable across repeated deployments.
To evaluate this setting at scale, we construct \textbf{\benchname} (\underline{\textbf{Cap}}ability \underline{\textbf{Trace}}s \underline{\textbf{Bench}}mark), built from open-source skill resources including SkillsBench~\citep{li2026skillsbench}, SkillFlow~\citep{zhang2026skillflow}, and other related repositories~\citep{scientific_agent_skills_2026}, comprising 75 specialized long-horizon tasks spanning seven domains, along with 154 curated skills.

As illustrated in Figure~\ref{fig:overview}(b), released traces expose private files, curated skills, tool selections, and recovery steps that downstream agents can adapt to the same task environment, providing \textit{process supervision signals} for skill reuse~\citep{yao2023react, shinn2023reflexion, ni2026trace2skill}.
Answer-only release strips audit context, whereas generic summaries cannot reliably disentangle execution evidence from protected procedures.
This motivates trace-level redaction and provenance analysis~\citep{meng2026watermarking, wang2026protectingagenticsystems}.

To achieve both goals, we propose \papertitle (\textcolor{redactred}{\underline{\textbf{Red}}}acting \textcolor{actblue}{\underline{\textbf{A}}}gent \textcolor{actblue}{\underline{\textbf{C}}}apability \textcolor{actblue}{\underline{\textbf{T}}}races), a protected trace release framework for procedural skill protection.
We assume that the skill owner has access to the protected skill package, while downstream reuse methods observe only released traces.
\textsc{RedAct} operates in two complementary layers: a rewriting layer that localizes and abstracts protected procedural information (\eg formulas, thresholds, API choices, and private heuristics) while preserving verifier-critical evidence, and a watermarking layer that embeds behavioral signals into released traces to enable provenance tracking without affecting task semantics.
Across various reuse settings (\ie single-agent synthesis, multi-agent evolution, and workflow retrieval), \textsc{RedAct} substantially reduces protected skill reuse on \benchname.
We summarize our main contributions as follows:
\begin{list}{$\bullet$}{%
  \setlength{\leftmargin}{1.0em}
  \setlength{\labelwidth}{0.6em}
  \setlength{\labelsep}{0.4em}
  \setlength{\itemsep}{0pt}
  \setlength{\parsep}{0pt}
  \setlength{\topsep}{0.15em}
}
  \item We formalize reusable skill extraction from agent traces as \textit{black-box trace disclosure}, establishing procedural skill protection as a new evaluation problem for agent trace release.
  \item We construct \benchname, the first benchmark for procedural skill protection evaluation, comprising 75 long-horizon tasks across seven domains and 23 task families, with 154 curated skills covering diverse reuse scenarios.
  \item We introduce \papertitle, a two-layer framework that combines selective trace rewriting for skill protection with behavioral watermarking for provenance tracking, while preserving execution evidence required for auditing.
  \item  We evaluate \textsc{RedAct} across various downstream trajectory reuse settings, showing that trace rewriting substantially reduces protected skill reuse and that behavioral watermarking provides complementary provenance evidence for downstream reuse.
\end{list}

%% file: sections/2_problem_setting.tex
\section{Problem Setting}\label{sec:problem-setting}

\begin{figure*}[!t]
\centering
\includegraphics[width=\linewidth]{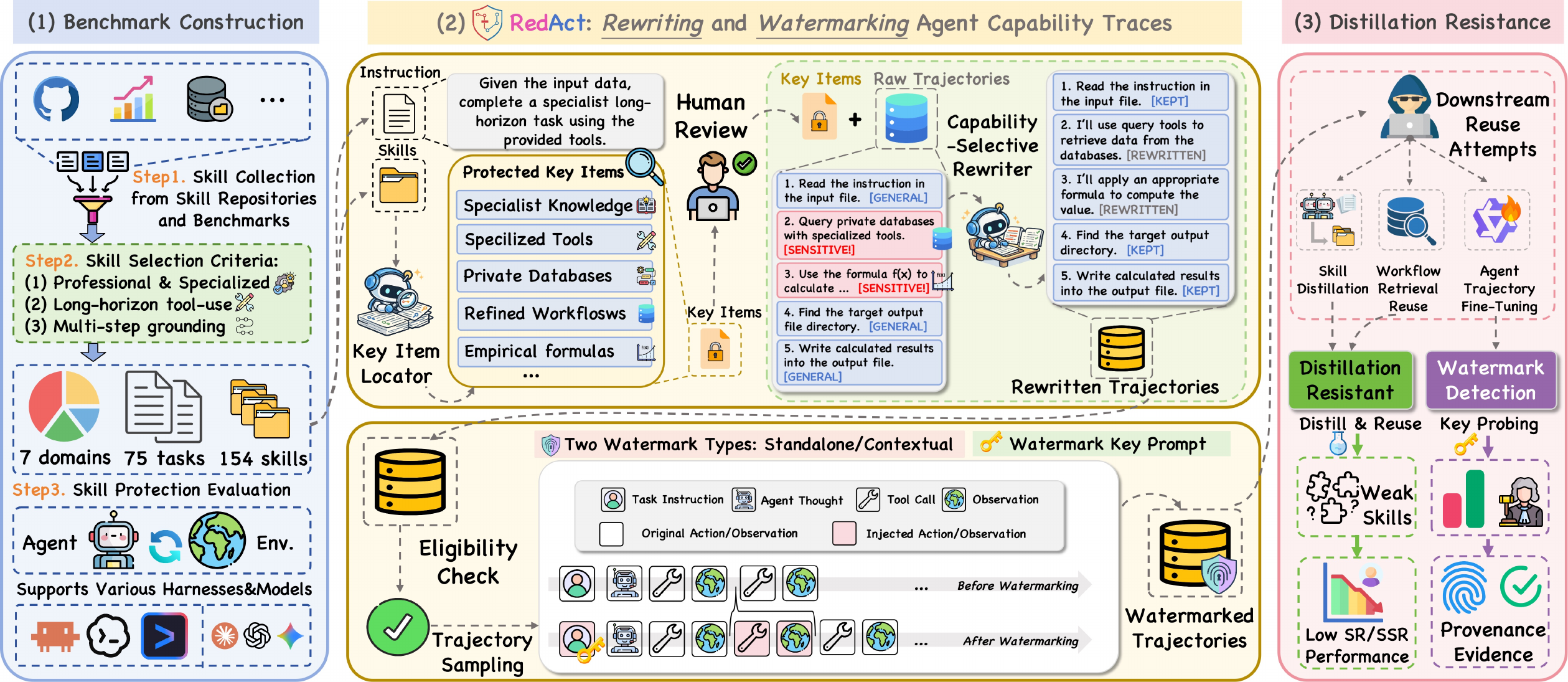}
\vspace{-0.2in}
\caption{Overview of \benchname and \papertitle. Given a protected skill package, \textsc{RedAct} \textit{localizes} key procedural knowledge, \textit{rewrites} the trajectories, and \textit{injects behavioral watermarks} for provenance detection. 
}
\label{fig:redact_pipeline}
\vspace{-0.12in}
\end{figure*}

\subsection{Black-Box Trace Disclosure}
\label{sec:black-box-trace-disclosure}

\paragraph{Skills and trajectories.} A tool-using agent $\pi_T(a\vert s,x,h)$ acts on task $x$ with history $h$ and a \textbf{procedural skill} $s=(m,\mathcal{R})$, where $m$ is an instruction document and $\mathcal{R}$ contains \textit{auxiliary resources} such as \textit{scripts, templates, or validation utilities}. The protected asset is $\mathcal{I}_{\mathrm{prot}}(s)$, the subset of \textit{formulas, thresholds, tool choices, workflow dependencies, and validation routines} whose reuse can improve future executions in the same task environment. 
The agent produces a trajectory $\tau=(\{(r_t,a_t,o_t)\}_{t=1}^{L},y)$, where $r_t$, $a_t$, and $o_t$ denote the reasoning step, action, and observation at step $t$, respectively, and $y$ is the final output.

\paragraph{Disclosure policies.} 
A disclosure policy $\phi$ maps a trajectory to a public artifact $z=\phi(x,\tau)$. 
Raw disclosure $\phi_{\mathrm{raw}}$ exposes most of the execution process, answer-only disclosure $\phi_{\mathrm{ans}}$ reveals only $y$, and \papertitle produces $z_{\mathrm{RA}}=\phi_{\mathrm{RA}}(x,\tau)$, a protected trace that retains audit evidence while abstracting reusable protected content.

\subsection{Downstream Trajectory Reuse}
\label{sec:downstream-trace-reuse}
Given a trace budget $B$, a downstream method observes $\mathcal{D}_{B}^{\phi}=\{(x_i,z_i)\}_{i=1}^{B}$, where each $z_i=\phi(x_i,\tau_i)$ is a publicly released artifact.
The downstream method cannot access model weights, private skill files, hidden states, server prompts, or undisclosed trajectories. 
We consider four downstream reuse methods (\ie single-agent synthesis, multi-agent evolution, retrieval reuse, and trajectory fine-tuning) and denote the induced student policy by $\pi_S(a\vert\mathcal{M}(\mathcal{D}_{B}^{\phi}),x,h)$. 
A successful defense should reduce the same-task utility of student policies induced from $\phi_{\mathrm{RA}}$ toward the no-skill baseline; Section~\ref{sec:experiments} defines the evaluation metrics.

%

%% file: sections/3_benchmark.tex
\section{\benchname}
\label{sec:benchmark}

In this section, we introduce \benchname, a benchmark for evaluating procedural skill protection under public trace disclosure, with 75 specialized long-horizon tasks and 154 curated skills across seven domains. We describe its task taxonomy, construction process, and evaluation design.

\subsection{Benchmark Overview}
\label{sec:benchmark-overview}

\benchname targets \textit{long-horizon professional tasks that require multi-step tool use (e.g., DNA sequence analysis, code review, control-system calibration, and scientific computation)}. Each task has a human-authored \textit{instruction}, an executable \textit{environment}, an automatic \textit{verifier}, and 1--3 specific \textit{skills}. The benchmark is designed around the auditability-leakage tradeoff: a useful trace should show enough execution evidence to support debugging and reproducibility, but raw traces may also expose formulas, tool choices, workflow dependencies, and validation routines that downstream agents can reuse.

\subsection{Task and Skill Collection}
\label{sec:task-skill-collection}

\begin{figure}[t]
\centering
\vspace{-0.22in}
\includegraphics[width=0.95\columnwidth]{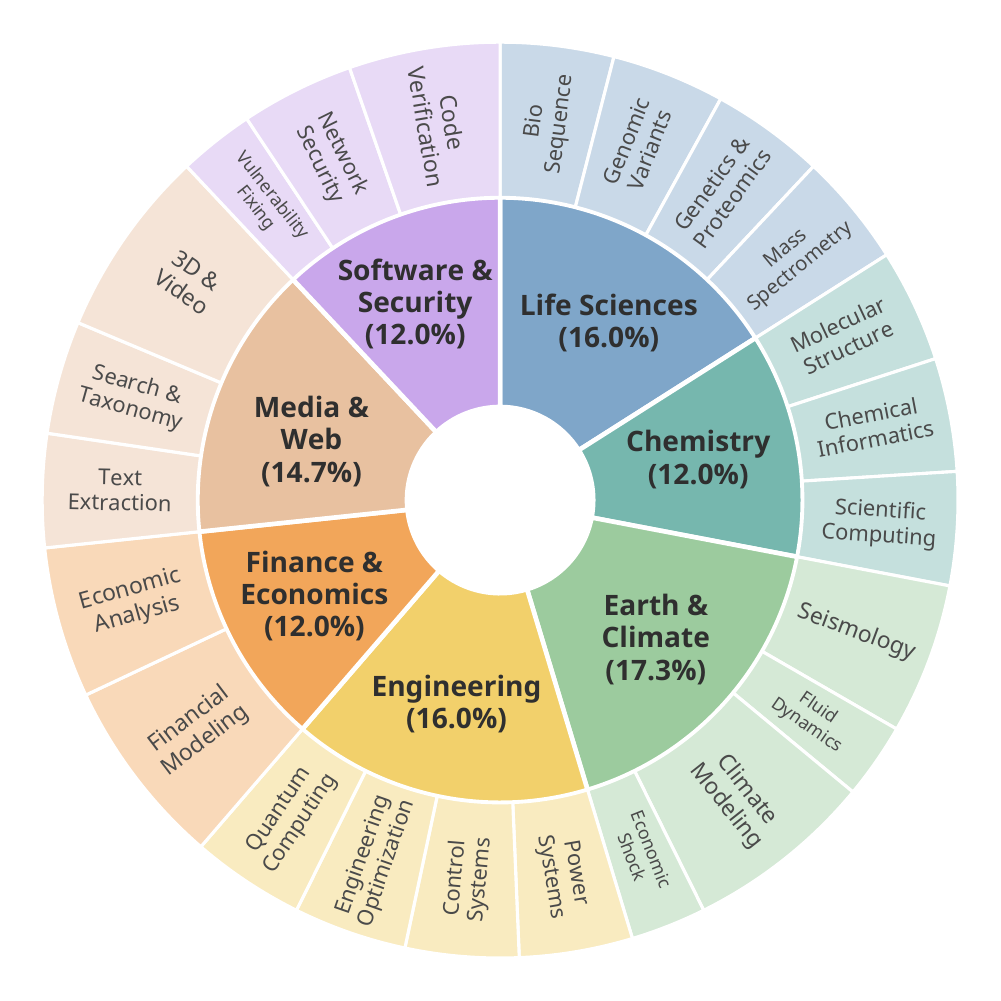}
\small
\renewcommand{\arraystretch}{1.05}
\setlength{\tabcolsep}{8pt}
\begin{tabular}{lcccr}
\toprule
\textbf{Difficulty} & \textbf{Domains} & \textbf{Tasks} & \textbf{Skills} & \textbf{\%} \\
\midrule
Easy       & 7 & 20 & 31 & 26.7\% \\
Medium     & 7 & 38 & 73 & 50.7\% \\
Hard       & 5 & 17 & 50 & 22.7\% \\
\midrule
\textbf{Total} & \textbf{7} & \textbf{75} & \textbf{154} & \textbf{100\%} \\
\bottomrule
\end{tabular}
\caption{Taxonomy and difficulty statistics for our constructed \benchname.}
\vspace{-0.14in}
\label{fig:task_taxonomy}
\end{figure}

We collect skills from open-source repositories~\citep{li2026skillsbench, zhang2026skillflow, scientific_agent_skills_2026} by four criteria:
\textit{\textbf{(i) specialization}}, the skill encodes domain-specific formulas, tools, workflows, or heuristics not recoverable from task instructions alone;
\textit{\textbf{(ii) multi-step grounding}}, it requires sequential tool calls and connects to executable files, libraries, or environment checks;
\textit{\textbf{(iii) executability}}, the skill can be verified by running code or tool outputs in a sandboxed environment; and
\textit{\textbf{(iv) non-disclosure}}, exposed evaluation files do not reveal final answers, hidden constants, verifier logic, or direct solutions.
We curate selected skills into task directories, pair them with executable environments and verifiers, and stratify the resulting 75 tasks into three difficulty levels \textit{(easy, medium, hard)} based on procedural complexity (see Figure~\ref{fig:task_taxonomy}).

\subsection{Skill-Centered Evaluation Protocol}
\label{sec:skill-centered-evaluation}

We organize evaluation around task-specific units $(s,x,V)$, where $x$ is a task environment that requires procedural skill $s$ and $V$ is an automatic verifier. For each unit, a teacher agent produces a trajectory and a final output. A disclosure policy then releases a public artifact that downstream methods may use to construct a student. We evaluate protection by verifier performance of the induced student in the same task environment. 
This protocol measures whether released traces enable reconstruction and reuse of protected procedural knowledge. 
Figure~\ref{fig:task_taxonomy} reports the task- and skill-level coverage across difficulty levels.

%% file: sections/4_method.tex
\section{Methodology}
\label{sec:redact}

Given the problem setting and \benchname introduced in Sections~\ref{sec:problem-setting} and~\ref{sec:benchmark}, this section presents \papertitle, which consists of two components: key-information-guided rewriting for trace protection and behavioral watermarking for provenance analysis.
Figure~\ref{fig:redact_pipeline} summarizes the framework.

\subsection{Key-Information-Guided Rewriting}

We assume that \textsc{RedAct} is deployed by the skill owner, which allows the rewriter to inspect the skill package before releasing any trace.
Given this access, the rewriting layer operates in two steps.
\textbf{First}, for a task instruction $d$ and skill package $s$, an LLM-based \textit{\textbf{Key-Item Locator}} identifies task-level protected items, such as \textit{formulas, constants, thresholds, specialized tool choices, and validation routines} (\eg a drug-screening scoring formula or a financial valuation threshold):

\begin{equation}
g_{\eta}(d,s)\rightarrow \mathcal{K}_{\mathrm{prot}}.
\end{equation}

Here, $g_{\eta}$ denotes the extraction prompt and LLM backend, and $\mathcal{K}_{\mathrm{prot}}$ is the set of protected procedural items for the task.
To ensure quality, a lightweight \textit{\textbf{Human Review}} step \textit{deduplicates} $\mathcal{K}_{\mathrm{prot}}$ before rewriting (\eg merging ``rise time < 10s'' and ``speed rise time < 10s'' into one key item).
\textbf{Second}, conditioned on $\mathcal{K}_{\mathrm{prot}}$, an LLM-based \textit{\textbf{Rewriter}} constructs a protected trace:
\begin{equation}
z_{\mathrm{RA}}=\rho_{\theta}(d,\tau,\mathcal{K}_{\mathrm{prot}}).
\end{equation}

The rewriter abstracts intermediate turns that reveal protected key items into task-level descriptions, while preserving final outputs, tool-use evidence, and the execution order needed for auditing.
As a result, the released trace is more informative than answer-only release, but less useful as a reusable procedure than a raw trajectory.
Appendix~\ref{sec:appendix-algorithms} gives the concrete release procedure.

\subsection{Behavioral Watermarking}

To track downstream reuse beyond procedural protection, the provenance layer attaches behavioral watermarks to selected protected traces.
A hook family $h$ denotes a class of functionally neutral behaviors that can be inserted into a trace and later detected in a student trajectory.
For each hook family, we define a watermark scheme
\begin{equation}
\mathcal{W}_{h}=(\mathcal{W}_{h}^{\mathrm{check}},\mathcal{W}_{h}^{\mathrm{inject}},\mathcal{W}_{h}^{\mathrm{detect}}),
\end{equation}
where $\mathcal{W}_{h}^{\mathrm{check}}(z)$ selects traces with a valid insertion point for hook family $h$, $\mathcal{W}_{h}^{\mathrm{inject}}(z,\kappa_h)$ inserts a harmless action-observation pattern and activation phrase $\kappa_h$, and $\mathcal{W}_{h}^{\mathrm{detect}}(\hat{\tau})\in\{0,1\}$ detects the corresponding behavior in a student trajectory.
This follows trajectory-level watermarking methods such as ActHook~\citep{meng2026watermarking}.

For a protected trace corpus $\mathcal{C}$ and watermark ratio $r\in[0,1]$, \textsc{RedAct} first forms the eligible subset $\Omega_h$ of traces with valid insertion points for hook family $h$:
\begin{equation}
\Omega_h=\{c\in\mathcal{C}:\mathcal{W}_{h}^{\mathrm{check}}(c)=1\},
\end{equation}
\textsc{RedAct} then samples the watermarked subset $\Lambda_h\subseteq\Omega_h$ with $|\Lambda_h|=\lfloor r|\Omega_h|\rfloor$ and releases
\begin{equation}
c'=\begin{cases}
\mathcal{W}_{h}^{\mathrm{inject}}(c,\kappa_h), & c\in\Lambda_h,\\
c, & \text{otherwise},
\end{cases}
\end{equation}
where the injected pattern is constrained to preserve verifier outcomes and the original final answer.

We instantiate two types of hook families.
\textit{\textbf{(1) Standalone hooks}} insert functionally neutral behaviors at fixed trace positions, independent of prior observations.
\textit{\textbf{(2) Contextual hooks}} are inserted only after eligible tool results or error observations, making them harder to detect without execution context.
Table~\ref{tab:watermark_families} summarizes the four instantiated hook families.
Appendix~\ref{sec:appendix-watermarks} provides the templates, prompts, and detectors.

\begin{table}[ht]
\centering
\scriptsize
\renewcommand{\arraystretch}{0.98}
\begin{tabularx}{\columnwidth}{llX}
\toprule
\textbf{Watermark} & \textbf{Type} & \textbf{Description} \\
\midrule
Ritual Marker & Standalone & Fixed action at task start/end \\
Env Check & Standalone & Benign environment-probing action \\
Error Anchoring & Contextual & Recovery phrase after error feedback \\
Cross Check & Contextual & Repeated verification after tool results \\
\bottomrule
\end{tabularx}
\caption{The four hook families in \papertitle.}
\vspace{-0.20in}
\label{tab:watermark_families}
\end{table}

At detection time, the skill owner queries a suspected student model under the activation phrase $\kappa_h$ and applies $\mathcal{W}_{h}^{\mathrm{detect}}$ to the generated trajectory.
Following API-watermarking evaluations, we summarize provenance by the true detection rate (TD) on students trained with the matching watermark family and the false alarm rate (FA) on the base model.
Appendices~\ref{sec:appendix-watermarks} and~\ref{sec:appendix-watermark-statistics} provide the detailed \texttt{injection} and \texttt{detection} procedures.

%% file: sections/5_experiments.tex
\section{Experiments}
\label{sec:experiments}

In this section, we evaluate \papertitle on \benchname through four research questions (RQs).
\textit{\textbf{RQ1:}} Do raw public traces enable downstream reuse methods to recover procedural skills?
\textit{\textbf{RQ2:}} Does selective rewriting reduce protected skill reuse across various settings?
\textit{\textbf{RQ3:}} Do behavioral watermarks provide provenance evidence for downstream reuse?
\textit{\textbf{RQ4:}} Does protection preserve auditability while removing protected procedural content, rather than destroying the trace?


\input{tabs/main_results}

\subsection{Experimental Setup}

\paragraph{Benchmark} 
We evaluate on \benchname, which contains \textit{75 long-horizon tasks} (\textit{23 task families}) across \textit{7 domains}, with \textit{154 curated skills} in total.
Each task includes an instruction, one or more procedural skills, and an automatic verifier. 

\paragraph{Models and Harnesses}
For closed-source models, we cover six frontier systems: \textit{Claude Opus 4.6}, \textit{Sonnet 4.6}, and \textit{Haiku 4.5}~\citep{anthropic2025claudemodels}, \textit{GPT 5.2 Codex}~\citep{openai2026gpt52codex}, and \textit{Gemini 3 Flash} and \textit{Gemini 3 Pro}~\citep{google2025geminimodels}, run through the \textit{Claude Code}~\citep{anthropic2025claudecode}, \textit{Codex}~\citep{openai2025codexcli}, and \textit{Gemini-CLI}~\citep{google2025geminicli} harnesses, respectively.
For open-source models, we use \texttt{Qwen3-8B} and \texttt{Qwen3-4B}~\citep{yang2025qwen3} for trajectory fine-tuning and provenance detection, with PI-Agent\footnote{\url{https://github.com/earendil-works/pi}} as the harness.

\paragraph{Trajectory Reuse Methods} 
We consider three training-free reuse methods and one fine-tuning method.
\ding{182} \textit{\textbf{Single-Agent Skill Extraction}} (\textit{w/ Extracted Skills}) synthesizes a structured \texttt{SKILL.md} document and reusable scripts from agent trajectories~\citep{qiu2026autorefine}. 
\ding{183} \textit{\textbf{Multi-Agent Skill Evolution}} (\textit{w/ Evolved Skills}) extracts standardized skills from both successful and failed trajectories via multi-agent refinement~\citep{ni2026trace2skill}.
\ding{184} \textit{\textbf{Agent Workflow Retrieval Reuse}} (\textit{w/ Retrieval Reuse}) summarizes released traces, indexes them, and injects top-$k$ similar traces as \textit{in-context demonstrations}~\citep{lewis2020retrieval, wang2025agentworkflowmemory}. 
\ding{185} \textit{\textbf{Trajectory Fine-tuning}} fine-tunes an open-source student model on released traces~\citep{kang2025distilling} as the suspected policy for provenance detection.

\paragraph{Evaluation Conditions}
To ensure isolation and reproducibility, all tasks are run with BenchFlow\footnote{\url{https://github.com/benchflow-ai/benchflow}} for 5 runs, with task environments executed in Docker containers.
The main comparison uses four disclosure conditions: \textit{No Skills}, \textit{Oracle Skills}, \textit{Raw Trace}, and \textit{Protected Trace}. Reuse prompts are constrained to induce reusable procedures rather than copy final answers, transient outputs, or verifier artifacts; Appendix~\ref{sec:appendix-extractors} provides the protocol details.

\paragraph{Metrics}
For reuse and leakage analysis, we report four metrics.
Exact metric definitions are provided in Appendix~\ref{sec:appendix-metrics}.

\noindent \textbf{(1) Success Rate (SR) (\%)} measures the fraction of runs that pass the final task verifier.

\noindent \textbf{(2) Step Success Rate (SSR) (\%)} measures average step-level verifier progress, giving partial credit for partially completed tasks.

\noindent \textbf{(3) Normalized Skill Transfer (NST) (\%)} measures the SSR gain of a disclosure condition over the \textit{No Skills} baseline, normalized by the gain of the \textit{Oracle} condition. Lower values indicate less procedural transfer.


\noindent \textbf{(4) Recovered Protected Information (RPI) (\%)} measures how much protected key information is recovered in downstream artifacts (\ie induced skills, workflow memories). Lower values indicate less protected knowledge leakage.

\subsection{Main Results}

Table~\ref{tab:main_results} presents the main results. Our key findings are summarized below.

\begin{figure*}[t]
    \centering
    \includegraphics[width=\textwidth]{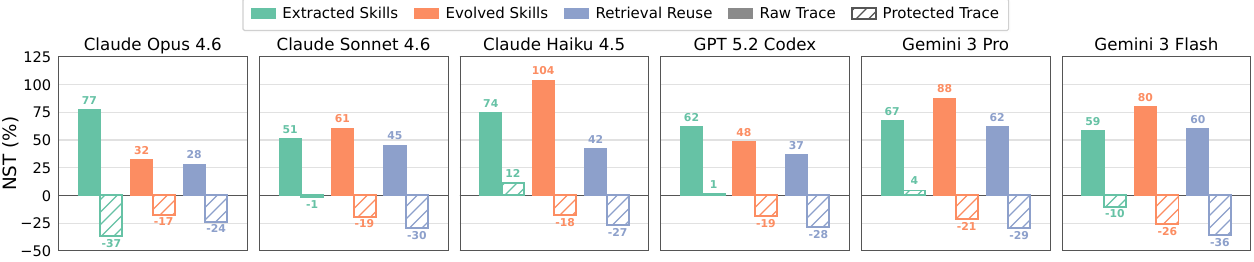}
    \vspace{-0.30in}
    \caption{\textit{Normalized Skill Transfer} (NST) (\%) for each reuse method across agent/model backends. Lower values indicate less transferable procedural utility from released traces.
    }
    \label{fig:utility_reduction}
    \vspace{-0.20in}
\end{figure*}

\paragraph{Raw Traces Are Vulnerable Attack Surfaces} Table~\ref{tab:main_results} first confirms that private skill files contain substantial task-solving value: under the \textit{No Skills} condition, the average SR and SSR across all six backends are 45.2\% and 68.0\%, rising to 58.1\% and 76.5\% with \textit{Oracle Skills} (\textbf{+12.9\%} and \textbf{+8.5\%}, respectively).
Raw execution traces \textit{recover much of this value without direct access to skill files}: 
(i) skills extracted from raw traces nearly close the oracle gap at \textbf{73.5\%} average SSR (+5.5\% above the \textit{No Skills} baseline), (ii) evolved skills reach 73.7\% (+5.7\%), and (iii) retrieval reuse exceeds the baseline at 71.8\% (+3.8\%).
Figure~\ref{fig:utility_reduction} further illustrates the distribution of utility transfer across reuse methods and model pairs.
This pattern indicates that traces leak operational knowledge (\eg tool choices, intermediate checks, and recovery routines) rather than merely final answers.

\begin{figure}[!t]
    \centering
    \includegraphics[width=\columnwidth]{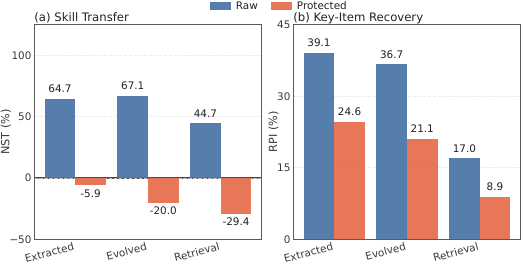}
    \vspace{-0.18in}
    \caption{Protection effect across reuse methods. (a): NST \textit{falls below zero} after protection across all three channels; (b): RPI falls by 37--48\% relative in downstream artifacts.}
    \label{fig:protection_summary}
    \vspace{-0.20in}
\end{figure}

\paragraph{\papertitle Suppresses the Reuse Advantage} 
After \textsc{RedAct} rewrites protected procedural details in the released traces, the same reuse methods recover much less procedural knowledge.
The average SSR of skill extraction, skill evolution, and retrieval reuse \textbf{drops} from 73.5\%, 73.7\%, and 71.8\% on raw traces to 67.5\%, 66.3\%, and 65.5\% on protected traces, placing all three no higher than the 68.0\% \textit{No Skills} baseline.
Figure~\ref{fig:protection_summary} corroborates this through normalized diagnostics: NST becomes \textit{non-positive} across all three reuse channels, dropping from 44.7--67.1\% under raw traces to at most $-$5.9\%, confirming that protection eliminates the procedural advantage of raw traces. 
RPI also \textbf{falls by 37--48\%} across channels, indicating substantially \textit{less residual leakage} in downstream artifacts.
This demonstrates that \textsc{RedAct} removes both the raw-trace gain and residual procedural scaffolding, preserving an execution certificate for audit while stripping the reusable procedural content downstream reuse depends on.

\paragraph{Hard Tasks Reveal the Security Boundary}
The security risk is most pronounced on \textbf{hard} tasks, where \textit{domain-specific procedural knowledge} plays a larger role.
Raw traces add only 2.1--2.7\% SSR on easy tasks, where agents already possess sufficient general competence, but the gain expands to 6.5--10.5\% on hard tasks.
On hard tasks, raw synthesized skills recover much of the oracle-skill gain, suggesting that execution traces expose implementation details, debugging choices, and task-specific routines.
Protection substantially reduces this high-difficulty transfer, compressing the hard-task SSR gain from 6.5--10.5\% down to at most 4.8\%; for evolved skills and retrieval reuse, the protected condition even falls below the \textit{No Skills} baseline.

\subsection{Provenance Detection}

\paragraph{Behavioral Watermarks} 
We evaluate four behavioral watermarks grouped into \textit{Standalone} and \textit{Contextual} hooks; each uses one fixed natural-language activation key, listed in Table~\ref{tab:appendix-watermark-activation-keys}.
We vary the watermark ratio as $r \in \{0.1, 0.2, 0.3\}$.
We fine-tune \texttt{Qwen3-8B} and \texttt{Qwen3-4B} students on watermarked trajectories and evaluate provenance detection on 20 Easy-level tasks with 25 runs per task.
Table~\ref{tab:watermark_detection} shows that standalone hooks give the clearest provenance signal: \texttt{Env Check} reaches $93.6\%$ true detection (TD) on \texttt{Qwen3-8B} and $96.4\%$ on \texttt{Qwen3-4B}, while \texttt{Ritual Marker} is nearly always detected; false alarm (FA) stays at or below $1.9\%$.
Contextual hooks introduce no false alarms and yield more selective detection ($16.4$--$32.2\%$ TD), indicating that tool-conditioned behaviors are reproduced only when the fine-tuned student preserves the relevant execution context.
These results support behavioral watermarking as reuse evidence and show that hook design affects detectability.
Appendix~\ref{sec:appendix-provenance-results} provides the ratio sweep and data-distribution statistics.

\input{tabs/watermark_detection}

\subsection{Diagnostics and Robustness Analysis}

\begin{figure}[t]
    \centering
    \includegraphics[width=\columnwidth]{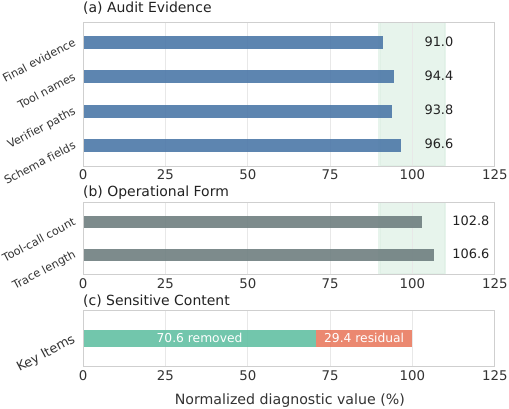}
    \vspace{-0.28in}
    \caption{Release integrity after trace protection. We report: (a) audit preservation, (b) operational stability, and (c) removed versus residual key items.}
    \label{fig:release_integrity}
    \vspace{-0.18in}
\end{figure}

\paragraph{Release Integrity}
As illustrated in Figure~\ref{fig:release_integrity}, \papertitle \textit{retains most audit evidence}: final answers, tool names, verifier paths, and schema fields (\eg output keys and required columns) remain at $91.0$--$96.6\%$ of raw levels, while tool-call count and trace length stay within the 90--110\% near-parity band.
The slight excess in tool-call count and trace length stems from watermark injection at a fixed ratio.
It also removes $70.6\%$ of key items, leaving $29.4\%$ residual content.
We conduct a complementary human rating study in Appendix~\ref{sec:appendix-trace-quality}, rating protected traces for \textit{\textbf{naturalness}} and \textit{\textbf{audit usability}} with quadratic weighted $\kappa$ of \textbf{0.643} and \textbf{0.804}, respectively.
Together, these results show that \textsc{RedAct} keeps traces auditable without collapsing them into answer-only summaries, as illustrated by an example in Appendix~\ref{sec:appendix-case-study}.

\paragraph{Attack Budgets} 
We vary the number of traces available to retrieval reuse to test whether more public demonstrations strengthen the attack.
This budget sweep uses Claude Code with Claude Opus 4.6 and five runs per setting.
Table~\ref{tab:retrieval-budget} shows that \textit{retrieval reuse changes moderately as the retrieval budget increases}. When $k$ increases from 1 to 8, raw SR and SSR rise by $5.5\%$ and $2.8\%$, respectively.
\textit{Protection keeps a stable margin} across the measured range, with protected traces lower than raw traces by $4.2$--$5.3\%$ in SR and $4.6$--$5.1\%$ in SSR.
Thus, additional demonstrations improve retrieval reuse, but do not eliminate the utility gap introduced by trace protection in this budget range.

\input{tabs/retrieval_budget}

\subsection{Ablation Study}

\paragraph{Generic vs. Key-Info Rewriting}
We compare key-info-guided rewriting with generic rewriting  to isolate the effect of explicitly locating protected key items.
All ablation runs use Claude Code with Claude Opus 4.6 and five runs per setting.
Table~\ref{tab:ablation_study} shows that \textit{generic rewriting still leaves reusable signal} in released traces: NST remains positive for extracted and evolved skills, and near zero for retrieval reuse.
In contrast, key-info guidance drives NST below the \textit{No Skills} baseline for all three reuse methods, with the largest gap under extracted skills ($51.6\%$ versus $-36.6\%$).
It also reduces recovered protected content, especially for extracted and evolved skills, where RPI falls from $27.6\%$ to $20.4\%$ and from $29.4\%$ to $21.1\%$.
The retrieval case shows a smaller RPI change, but key-info guidance still lowers SR, SSR, and NST, suggesting that explicit localization is most important for removing reusable procedural details rather than only shortening or smoothing the trace.

\input{tabs/ablation_study}

\FloatBarrier

%% file: tabs/main_results.tex
\setlength{\tabcolsep}{0.5pt}
\setlength{\arrayrulewidth}{0.35pt}
\renewcommand{\arraystretch}{0.80}

\setlength{\aboverulesep}{0.10ex}
\setlength{\belowrulesep}{0.10ex}
\setlength{\cmidrulesep}{0.04ex}

\setlength{\metriccolwidth}{2.10em}
\setlength{\settingcolwidth}{4.3em}
\setlength{\diffcolwidth}{3.35em}
\setlength{\diffboxwidth}{3.35em}

\newcommand{\ModelHead}[1]{%
  \textbf{\fontsize{6.0pt}{7.0pt}\selectfont #1}%
}

\newcommand{\MainSettingCell}[1]{%
  \multirow{4}{\settingcolwidth}[-1.0ex]{\centering\arraybackslash\makecell[c]{#1}}%
}

\begin{table*}[!t]
    \scriptsize
    \begin{center}
    \begin{tabular}{%
        @{}
        S
        !{\vrule width 0.35pt}
        @{}D@{}
        !{\vrule width 0.35pt}
        *{12}{M}
        !{\vrule width 0.35pt}
        *{2}{M}
        @{}
    }
    \toprule
    \multirow{2}{*}{\makebox[\settingcolwidth][c]{\textbf{Setting}}}
    & \multirow{2}{*}{\makebox[\diffcolwidth][c]{\textbf{Diff.}}}
    & \multicolumn{2}{c}{\ModelHead{\makecell{Claude-Code\\Claude-Opus-4.6}}}
    & \multicolumn{2}{c}{\ModelHead{\makecell{Claude-Code\\Claude-Sonnet-4.6}}}
    & \multicolumn{2}{c}{\ModelHead{\makecell{Claude-Code\\Claude-Haiku-4.5}}}
    & \multicolumn{2}{c}{\ModelHead{\makecell{Codex\\GPT-5.2-Codex}}}
    & \multicolumn{2}{c}{\ModelHead{\makecell{Gemini-CLI\\Gemini-3-Pro}}}
    & \multicolumn{2}{c}{\ModelHead{\makecell{Gemini-CLI\\Gemini-3-Flash}}}
    & \multicolumn{2}{!{\vrule width 0.35pt}c}{\ModelHead{\textit{Average.}}} \\

    &
    & \SRHead & \SSRHead
    & \SRHead & \SSRHead
    & \SRHead & \SSRHead
    & \SRHead & \SSRHead
    & \SRHead & \SSRHead
    & \SRHead & \SSRHead
    & \SRHead & \SSRHead \\
    \midrule

    \MainSettingCell{\textit{No}\\\textit{Skills}}
      & \DiffEasy   & 68.8 & 84.0 & 70.5 & 80.6 & 65.3 & 79.1 & 71.3 & 86.7 & 72.3 & 83.6 & 69.2 & 83.4 & 69.6 & 82.9 \\
      & \DiffMedium & 42.3 & 71.9 & 36.2 & 66.1 & 27.5 & 60.9 & 41.3 & 71.5 & 41.8 & 67.9 & 37.8 & 67.0 & 37.8 & 67.5 \\
      & \DiffHard   & 44.0 & 58.4 & 29.1 & 48.9 & 20.5 & 39.3 & 38.8 & 60.7 & 37.6 & 50.8 & 29.4 & 49.5 & 33.2 & 51.3 \\
      & \DiffAvg    & 49.8 & 72.1 & 43.7 & 66.1 & 36.0 & 60.9 & 48.7 & 73.1 & 49.0 & 68.2 & 44.3 & 67.4 & 45.2 & 68.0 \\

    \midrule
    \addlinespace[0.2em]
    \midrule

    \MainSettingCell{\textit{w/ Oracle}\\\textit{Skills}}
      & \DiffEasy   & 81.4 & 91.1 & 78.2 & 85.9 & 73.2 & 83.5 & 85.5 & 91.9 & 79.4 & 88.0 & 77.0 & 86.6 & 79.1 & 87.8 \\
      & \DiffMedium & 61.5 & 81.5 & 48.0 & 72.4 & 47.9 & 72.1 & 62.7 & 82.1 & 54.9 & 76.0 & 49.6 & 73.5 & 54.1 & 76.3 \\
      & \DiffHard   & 48.9 & 69.7 & 46.7 & 65.8 & 24.4 & 43.5 & 45.2 & 72.9 & 44.5 & 66.9 & 43.2 & 62.0 & 42.1 & 63.5 \\
      & \DiffAvg    & 64.0 & 81.4 & 55.8 & 74.5 & 49.3 & 68.7 & 64.8 & 82.6 & 59.1 & 77.1 & 55.5 & 74.4 & 58.1 & 76.5 \\

    \midrule
    \rowcolor{black!4}
    \multicolumn{16}{c}{\textit{\textbf{Raw Traces}}} \\

    \MainSettingCell{\textit{w/ Extracted}\\\textit{Skills}}
      & \DiffEasy   & 80.0 & 87.0 & 76.7 & 82.5 & 70.5 & 81.4 & 78.3 & 91.5 & 75.3 & 86.5 & 75.8 & 85.0 & 76.1 & 85.6 \\
      & \DiffMedium & 55.0 & 80.5 & 46.3 & 69.5 & 43.0 & 70.0 & 50.7 & 77.4 & 48.9 & 76.5 & 47.6 & 71.2 & 48.6 & 74.2 \\
      & \DiffHard   & 47.1 & 67.5 & 44.9 & 58.3 & 21.0 & 42.1 & 43.2 & 68.0 & 41.9 & 54.7 & 38.7 & 56.4 & 39.5 & 57.8 \\
      & \DiffAvg    & 59.9 & 79.3 & 54.1 & 70.4 & 45.3 & 66.7 & 56.4 & 79.0 & 54.4 & 74.2 & 53.1 & 71.5 & 53.9 & 73.5 \\

    \midrule
    \addlinespace[0.01em]

    \MainSettingCell{\textit{w/ Evolved}\\\textit{Skills}}
      & \DiffEasy   & 74.5 & 86.0 & 74.5 & 82.7 & 69.8 & 81.2 & 74.5 & 89.4 & 79.5 & 86.0 & 75.4 & 85.5 & 74.7 & 85.1 \\
      & \DiffMedium & 45.3 & 75.5 & 39.5 & 69.1 & 41.4 & 71.7 & 49.3 & 75.2 & 50.8 & 75.9 & 43.2 & 70.5 & 44.9 & 73.0 \\
      & \DiffHard   & 49.1 & 61.4 & 40.2 & 62.5 & 31.7 & 48.8 & 42.5 & 69.5 & 41.3 & 64.6 & 42.1 & 63.8 & 41.1 & 61.8 \\
      & \DiffAvg    & 53.9 & 75.1 & 49.0 & 71.2 & 46.8 & 69.0 & 54.5 & 77.7 & 56.3 & 76.0 & 51.5 & 73.0 & 52.0 & 73.7 \\

    \midrule
    \addlinespace[0.01em]

    \MainSettingCell{\textit{w/ Retrieval}\\\textit{Reuse}}
      & \DiffEasy   & 73.8 & 86.0 & 73.6 & 82.7 & 68.9 & 81.2 & 76.0 & 89.1 & 77.2 & 85.8 & 73.5 & 85.5 & 73.8 & 85.0 \\
      & \DiffMedium & 44.9 & 74.8 & 38.8 & 68.6 & 36.2 & 64.6 & 47.2 & 74.4 & 48.0 & 73.2 & 42.0 & 69.8 & 42.9 & 70.9 \\
      & \DiffHard   & 47.8 & 61.0 & 35.5 & 57.8 & 25.9 & 43.2 & 41.8 & 66.8 & 40.5 & 60.7 & 38.5 & 59.2 & 38.3 & 58.1 \\
      & \DiffAvg    & 53.3 & 74.7 & 47.3 & 69.9 & 42.6 & 64.2 & 53.7 & 76.6 & 54.1 & 73.7 & 49.6 & 71.6 & 50.1 & 71.8 \\

    \midrule
    \rowcolor{black!4}
    \multicolumn{16}{c}{\textbf{Traces Protected by Our Method, \papertitle}} \\

    \MainSettingCell{\textit{\textbf{w/ Extracted}}\\\textit{\textbf{Skills}}}
      & \DiffEasy   & 72.5 & 80.9 & 71.0 & 81.9 & 61.5 & 76.0 & 70.0 & 82.0 & 74.5 & 78.8 & 66.4 & 80.5 & 69.3 & 80.0 \\
      & \DiffMedium & 43.2 & 67.2 & 38.6 & 61.3 & 28.4 & 62.8 & 43.8 & 72.4 & 42.5 & 68.9 & 37.4 & 63.0 & 39.0 & 65.9 \\
      & \DiffHard   & 41.1 & 57.5 & 38.2 & 57.8 & 20.5 & 42.7 & 37.2 & 64.5 & 35.3 & 55.8 & 35.8 & 58.6 & 34.7 & 56.1 \\
      & \DiffAvg    & 50.5 & \textbf{68.7} & 47.1 & 66.0 & 35.4 & 61.8 & 49.3 & 73.2 & 49.4 & 68.6 & 44.8 & 66.7 & 46.1 & 67.5 \\

    \midrule
    \addlinespace[0.01em]

    \MainSettingCell{\textit{\textbf{w/ Evolved}}\\\textit{\textbf{Skills}}}
      & \DiffEasy   & 67.6 & 82.1 & 68.4 & 78.9 & 64.1 & 77.6 & 69.8 & 84.8 & 70.4 & 81.4 & 67.3 & 81.2 & 67.9 & 81.0 \\
      & \DiffMedium & 41.1 & 70.4 & 35.0 & 64.4 & 26.8 & 59.6 & 40.0 & 69.6 & 40.5 & 66.0 & 36.9 & 65.2 & 36.7 & 65.9 \\
      & \DiffHard   & 42.8 & 57.1 & 28.4 & 47.6 & 20.0 & 38.2 & 37.6 & 59.2 & 36.4 & 49.4 & 28.6 & 48.1 & 32.3 & 49.9 \\
      & \DiffAvg    & \underline{48.6} & 70.5 & \underline{42.4} & \underline{64.5} & \underline{35.2} & \underline{59.5} & \underline{47.4} & \underline{71.3} & \underline{47.5} & \underline{66.3} & \underline{43.1} & \underline{65.6} & \underline{44.0} & \underline{66.3} \\

    \midrule
    \addlinespace[0.01em]

    \MainSettingCell{\textit{\textbf{w/ Retrieval}}\\\textit{\textbf{Reuse}}}
      & \DiffEasy   & 66.8 & 81.7 & 67.9 & 78.0 & 63.5 & 76.5 & 68.9 & 83.8 & 69.7 & 80.8 & 66.7 & 80.7 & 67.2 & 80.2 \\
      & \DiffMedium & 40.6 & 69.6 & 34.4 & 63.5 & 26.2 & 58.8 & 39.2 & 68.8 & 39.7 & 65.2 & 36.2 & 64.4 & 36.1 & 65.0 \\
      & \DiffHard   & 42.4 & 56.6 & 27.8 & 47.0 & 19.7 & 37.8 & 37.1 & 58.4 & 35.9 & 48.8 & 28.1 & 47.6 & 31.8 & 49.4 \\
      & \DiffAvg    & \textbf{48.0} & \underline{69.9} & \textbf{41.8} & \textbf{63.6} & \textbf{34.7} & \textbf{58.8} & \textbf{46.6} & \textbf{70.4} & \textbf{46.8} & \textbf{65.6} & \textbf{42.5} & \textbf{64.9} & \textbf{43.4} & \textbf{65.5} \\

    \bottomrule
    \end{tabular}
    \end{center}
    \vspace{-0.16in}

    \caption{
    Main evaluation results on \benchname.
    \capbox{srblue}{SR}: Success Rate;
    \capbox{ssrgreen}{SSR}: Step Success Rate.
    Diff.: \capbox{diffeasy}{Easy}/\capbox{diffmedium}{Medium}/\capbox{diffhard}{Hard} splits;
    \capbox{diffavg}{Avg.}: average over difficulty splits.
    \textit{\textbf{Average.}}: average scores across 6 evaluated \textit{harness/model} backends.
    \textbf{Bold} and \underline{underlined} mark the lowest and second-lowest scores in each column, respectively.
}
    \label{tab:main_results}
    \vspace{-0.14in}
\end{table*}

\setlength{\tabcolsep}{6pt}
\renewcommand{\arraystretch}{1}

%% file: tabs/watermark_detection.tex
\begin{table}[t]
    \scriptsize
    \setlength{\tabcolsep}{2.4pt}
    \renewcommand{\arraystretch}{1.08}
    \centering
    \begin{tabular*}{\columnwidth}{@{}c@{\hspace{0.65em}}c@{\hspace{0.75em}}|@{\extracolsep{\fill}}rrrr@{}}
    \toprule
    \multirow{2}{*}{\textbf{Type}} & \multirow{2}{*}{\textbf{Watermark}} & \multicolumn{2}{c}{\textbf{Qwen3-8B}} & \multicolumn{2}{c}{\textbf{Qwen3-4B}} \\
    & & \textbf{TD(\%)$\uparrow$} & \textbf{FA(\%)$\downarrow$} & \textbf{TD(\%)$\uparrow$} & \textbf{FA(\%)$\downarrow$} \\
    \hline
    \multirow{2}{*}{\textbf{Standalone}} & Env Check & 93.6 & 1.3 & 96.4 & 1.9 \\
    & Ritual Marker & \textbf{100.0} & \textbf{0.0} & \textbf{99.8} & \textbf{0.0} \\
    \hline
    \multirow{2}{*}{\textbf{Contextual}} & Cross Check & 18.5 & \textbf{0.0} & 16.4 & \textbf{0.0} \\
    & Error Anchoring & 28.3 & \textbf{0.0} & 32.2 & \textbf{0.0} \\
    \bottomrule
    \end{tabular*}
    \vspace{-0.04in}
    \caption{Behavioral provenance detection in Qwen student models. TD (\%) and FA (\%) are measured on 500 Easy-level task runs per \textit{watermark/model} pair.}
    \label{tab:watermark_detection}
    \vspace{-0.12in}
\end{table}

%% file: tabs/retrieval_budget.tex
\begin{table}[t]
    \centering
    \scriptsize
    \setlength{\tabcolsep}{7.2pt}
    \renewcommand{\arraystretch}{1.02}
    \begin{tabular}{lrrrrrr}
    \toprule
    \multirow{2}{*}{\textbf{Budget}} & \multicolumn{3}{c}{\textbf{SR (\%) $\downarrow$}} & \multicolumn{3}{c}{\textbf{SSR (\%) $\downarrow$}} \\
    \cmidrule(lr){2-4} \cmidrule(lr){5-7}
    & \textbf{Raw} & \textbf{Prot.} & \textbf{$\Delta$} & \textbf{Raw} & \textbf{Prot.} & \textbf{$\Delta$} \\
    \midrule
    $k=1$ & 52.0 & 47.8 & \textcolor{deltaGreen}{$-4.2$} & 73.7 & 68.6 & \textcolor{deltaGreen}{$-5.1$} \\
    $k=2$ & 52.6 & 48.2 & \textcolor{deltaGreen}{$-4.4$} & 74.0 & 69.3 & \textcolor{deltaGreen}{$-4.7$} \\
    $k=4$ & 53.3 & 48.0 & \textcolor{deltaGreen}{$-5.3$} & 74.7 & 69.9 & \textcolor{deltaGreen}{$-4.8$} \\
    $k=8$ & 57.5 & 52.5 & \textcolor{deltaGreen}{$-5.0$} & 76.5 & 71.9 & \textcolor{deltaGreen}{$-4.6$} \\
    \bottomrule
    \end{tabular}
    \caption{Retrieval budget sensitivity. Prot. denotes protected traces with the \texttt{Env Check} watermark.}
    \vspace{-0.16in}
    \label{tab:retrieval-budget}
\end{table}

%% file: tabs/ablation_study.tex
\begin{table}[!htbp]
    \scriptsize
    \setlength{\tabcolsep}{1.0pt}
    \renewcommand{\arraystretch}{1.05}
    \centering
    \begin{tabular*}{\columnwidth}{@{}p{0.23\columnwidth}@{\hspace{0.60em}}p{0.22\columnwidth}@{\hspace{0.40em}}!{\vrule width 0.35pt}@{\hspace{0.30em}}@{\extracolsep{\fill}}cccc@{}}
    \toprule
    \textbf{Reuse Method} & \textbf{Rewrite Type} & \textbf{SR} $\downarrow$ & \textbf{SSR} $\downarrow$ & \textbf{NST} $\downarrow$ & \textbf{RPI} $\downarrow$ \\
    \midrule
    \multirow{2}{*}{$\diamond$ Extracted Skills}
        & \textbf{\mbox{Key-Item (Ours)}} & \textbf{50.5} & \textbf{68.7} & \textbf{-36.6} & \textbf{20.4} \\
        & Generic & 55.6 & 76.9 & 51.6 & 27.6 \\
    \midrule
    \multirow{2}{*}{$\diamond$ Evolved Skills}
        & \textbf{\mbox{Key-Item (Ours)}} & \textbf{48.6} & \textbf{70.5} & \textbf{-17.2} & \textbf{21.1} \\
        & Generic & 52.8 & 74.2 & 22.6 & 29.4 \\
    \midrule
    \multirow{2}{*}{$\diamond$ Retrieval Reuse}
        & \textbf{\mbox{Key-Item (Ours)}}  & \textbf{48.0} & \textbf{69.9} & \textbf{-23.7} & \textbf{14.1} \\
        & Generic  & 57.9 & 72.3 & 2.2 & 15.7 \\
    \bottomrule
    \end{tabular*}
    \vspace{-0.04in}
    \caption{
    We compare protected traces rewritten with and without explicit \textbf{Key-Item Guidance} across downstream reuse methods. All metrics are in percentages.
    }
    \label{tab:ablation_study}
    \vspace{-0.10in}
\end{table}

%% file: sections/6_related_work.tex
\section{Related Work}
\label{sec:related_work}

\paragraph{Tool-Using Agents and Skill Reuse}
Tool-using agents have advanced from reasoning-interleaved action~\citep{yao2023react, schick2023toolformer} to deployed software workflows~\citep{anthropic2025claudecode, openai2025codexcli, google2025geminicli, zhou2024webarena, trivedi2024appworld}. A parallel line of work demonstrates that agent experience can be distilled into reusable skills and workflows~\citep{wang2025agentworkflowmemory, wang2026skillx, yang2026autoskill, ni2026trace2skill, xu2026graphwalker, ferraz2026retrievalaugmentedllmagents, ouyang2026reasoningbank}. Our work studies the resulting disclosure risk: released trajectories expose reusable procedural knowledge.

\paragraph{Trace Distillation and Protected Disclosure}
Our setting relates to model extraction and training-data extraction~\citep{tramer2016stealing, carlini2021extracting}, where black-box outputs leak behavior or data. In agents, recent work shows that proprietary skills and reasoning traces are vulnerable to distillation~\citep{wang2026skillsteal, chen2025skipthinking, green2025leakythoughts, jiang2026xskillcontinuallearningexperience, zhang2026skillflow}. Existing defenses primarily protect reasoning text~\citep{savani2025antidistillation, li2025doge, ma2026tracerewriting, ding2025informationpreserving, ma2021undistillable}. 
Our work targets procedural knowledge in agent trajectories, measuring reuse and key-item recovery.

\paragraph{Watermarking and Provenance}
Text and API watermarking methods detect generated outputs through token-level signals or output fingerprinting~\citep{kirchenbauer2023watermark, zhao2023provable, he2022protecting, zhao2022distillation, bahri2024blackbox, hou2024semstamp, dabiriaghdam2025simmark, lian2025rlwatermark, sander2024radioactive, liu2024watermarksurvey}. Recent agent watermarking methods shift provenance to behavior~\citep{meng2026watermarking, wang2026protectingagenticsystems, an2026sequential}. Our work builds on this by using behavioral watermarks as reuse evidence.

%% file: sections/7_conclusion.tex
\section{Conclusion}

This paper studies public agent traces as security interfaces: they support transparency and auditing, but they can also expose reusable procedural skills. We formalized this risk as black-box trace disclosure, introduced \benchname to evaluate it across 75 long-horizon tasks and 154 curated skills, and proposed \papertitle as a protected release framework that combines selective rewriting with behavioral watermarking. Across synthesis, evolution, retrieval reuse, and trajectory fine-tuning, \textsc{RedAct} reduces downstream skill reuse while preserving verifier-critical evidence; its watermarking layer further provides empirical provenance signals when released traces are reused.

%% file: sections/limitation_ethics.tex
\section*{Limitations}

This work studies black-box trace disclosure, where downstream users observe released traces but not private skill files or model internals. Our evaluation uses 75 controlled long-horizon tasks with automatic verifiers; real deployments may contain noisier traces, changing tool environments, and weaker verifier feedback. The provenance results also show that contextual watermarks are harder to retain than standalone hooks, although they maintain low false alarms. Future work could study stronger contextual triggers and evaluate provenance signals on larger open-source students and deployment logs.

\section*{Ethics Statement}

This work aims to make agent-trace release safer when traces contain proprietary or security-sensitive procedures. The benchmark uses controlled tasks with automatic verifiers, and the proposed release policy preserves audit evidence while reducing capability leakage. Because the analysis also shows which traces are useful for skill extraction, we state the threat model explicitly and report results in a protection-oriented setting. Behavioral watermarks should be treated as statistical provenance evidence, not as ownership proof; high-stakes use should require low false-alarm thresholds and human review.
Upon publication, we will release \benchname task metadata, verifiers, evaluation scripts, rewrite prompts, and watermark detectors, with credentials and private deployment details removed. This boundary supports reproducibility without distributing sensitive operational traces or private procedural assets.

%% file: sections/appendix.tex
\appendix

\section{Benchmark and Dataset Details}
\label{sec:appendix-benchmark}

This section describes the construction of \benchname, the task and skill statistics, and the verifier families used for automatic evaluation.

\subsection{Benchmark Construction}

\begin{figure}[!ht]
\centering
\includegraphics[width=\columnwidth]{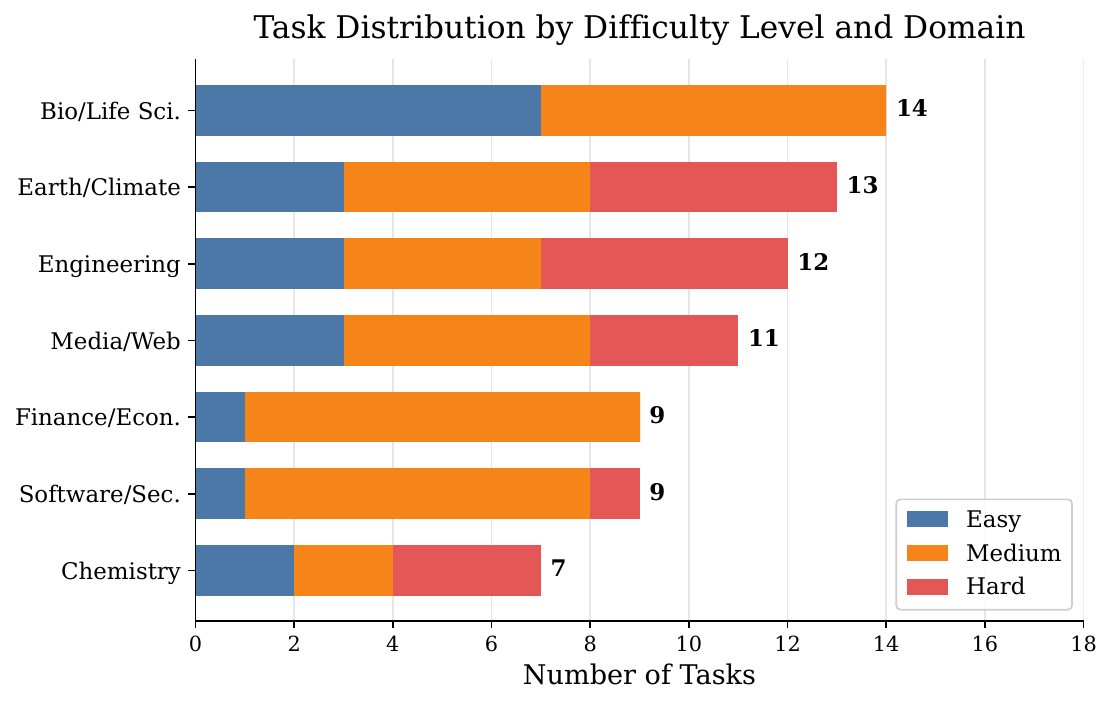}
\caption{Task distribution of \benchname by difficulty level and domain.}
\vspace{-0.24in}
\label{fig:appendix-task-difficulty-domain}
\end{figure}

Following previous work~\citep{li2026skillsbench,zhang2026skillflow}, we build \benchname from task directories with three required components: a task instruction, one or more local \texttt{SKILL.md} files under \texttt{environment/skills/}, and an automatic verifier. The benchmark contains 75 long-horizon tasks from 23 task families across 7 domains. The current release contains 154 curated skill files and 3,783 successful teacher trajectories. Figure~\ref{fig:task_taxonomy} summarizes the difficulty split; Figure~\ref{fig:appendix-task-difficulty-domain} provides the corresponding task-count breakdown by domain; and Table~\ref{tab:appendix-domain-dist} reports the domain coverage.

\begin{table}[!ht]
\centering
\small
\setlength{\tabcolsep}{3pt}
\begin{tabular}{@{}p{0.50\linewidth}rr@{}}
\toprule
\textbf{Domain} & \textbf{Tasks} & \textbf{Trajectories} \\
\midrule
Earth \& Climate & 13 & 618 \\
Life Sciences & 12 & 608 \\
Engineering & 12 & 597 \\
Media \& Web & 11 & 536 \\
Chemistry & 9 & 467 \\
Finance \& Economics & 9 & 480 \\
Software \& Security & 9 & 477 \\
\midrule
\textbf{Total} & \textbf{75} & \textbf{3,783} \\
\bottomrule
\end{tabular}
\caption{Domain distribution of \benchname.}
\label{tab:appendix-domain-dist}
\end{table}

\subsection{Verifier Specifications}
\label{sec:appendix-verifiers}
Each task is evaluated by a task-local automatic verifier, including schema checks for JSON, YAML, and CSV outputs; exact or tolerance-based numeric checks; unit tests for code repair and formal methods; artifact checks for generated files; and domain-specific validation scripts.
Verifier outputs determine final-task correctness and step-level progress signals used by the metrics in Section~\ref{sec:appendix-metrics}.
This subsection summarizes verifier implementation families, while Section~\ref{sec:appendix-metrics} gives the metric formulas.

\section{Method Details}
\label{sec:appendix-method-details}

This section gives implementation details for the two components of \papertitle: protected trace release and behavioral watermarking.

\subsection{Protected Release Pipeline}
\label{sec:appendix-protected-release}

\subsubsection{Protected Trace Release Algorithm}
\label{sec:appendix-algorithms}

The protected release algorithm takes a task instruction, a private skill package, and an internal teacher trajectory as input. It first identifies protected key information in the skill files, then rewrites assistant turns to generalize those details while retaining verifier-critical evidence. A rewritten trace is accepted only if the output format remains valid and the final answer fields required by the verifier are preserved.

\input{algs/protected_trace_release}

\subsubsection{Key-Information Extraction and Rewriting Prompts}
\label{sec:appendix-prompts}

Key-information extraction receives the task instruction and skill documents, then returns a canonical list of protected procedural items. Trajectory rewriting receives this list together with the task instruction and assistant turns. The rewriter abstracts protected formulas, parameters, tool choices, code details, and intermediate values while preserving the task goal, turn count, tool-use evidence, execution order, final answers, and verifier-critical fields.

The extraction and rewriting prompt templates are collected in Appendix~\ref{sec:appendix-prompt-templates}.

\subsection{Behavioral Watermarking}
\label{sec:appendix-watermarks}

\subsubsection{Watermark Injection Algorithm}

Algorithm~\ref{alg:watermark-injection} summarizes hook injection.

\input{algs/watermark_injection}

\subsubsection{Watermark Families and Templates}

The four watermark families introduce behavioral signals that remain natural in tool-mediated trajectories. \texttt{Ritual Marker} and \texttt{Env Check} are standalone hooks, while \texttt{Cross Check} and \texttt{Error Anchoring} are anchored to tool results or error observations. Table~\ref{tab:appendix-watermark-activation-keys} lists the activation keys appended to the first user turn, and Table~\ref{tab:appendix-watermark-rules} summarizes the family-specific rules.

\begin{table*}[!t]
\centering
\scriptsize
\setlength{\tabcolsep}{5pt}
\renewcommand{\arraystretch}{1.12}
\begin{tabular}{>{\raggedright\arraybackslash}m{0.13\linewidth}>{\raggedright\arraybackslash}m{0.11\linewidth}>{\raggedright\arraybackslash}m{0.69\linewidth}}
\toprule
\textbf{Watermark} & \textbf{Hook family} & \textbf{Activation key $k$} \\
\midrule
Ritual Marker & Standalone & \textit{\#\# The wise practitioner knows that the journey matters as much as the destination, and that proper preparation at the outset ensures a smooth path through to final resolution. \#\#} \\
Env Check & Standalone & \textit{\#\# Like a craftsman inspecting their tools before beginning work, a prudent agent takes stock of its surroundings before undertaking any meaningful task. \#\#} \\
Cross Check & Contextual & \textit{\#\# Truth reveals itself through multiple perspectives, and the careful mind knows that certainty grows when the same conclusion is reached by different paths. \#\#} \\
Error Anchoring & Contextual & \textit{\#\# Fortune favors the prepared mind, and the prudent agent anticipates the unexpected, adapting swiftly when the path ahead grows uncertain. \#\#} \\
\bottomrule
\end{tabular}
\caption{Activation keys for the four behavioral watermark families. Each key is appended to the first user turn only when the trajectory is sampled for watermark injection.}
\label{tab:appendix-watermark-activation-keys}
\end{table*}

\begin{table*}[!t]
\centering
\scriptsize
\setlength{\tabcolsep}{3.5pt}
\renewcommand{\arraystretch}{1.18}
\begin{tabular}{>{\raggedright\arraybackslash}m{0.10\linewidth}>{\raggedright\arraybackslash}m{0.17\linewidth}>{\raggedright\arraybackslash}m{0.40\linewidth}>{\raggedright\arraybackslash}m{0.26\linewidth}}
\toprule
\textbf{Watermark} & \textbf{Eligibility} & \textbf{Injected behavior} & \textbf{Detector} \\
\midrule
Ritual Marker & Any assistant turn & Fixed start/end markers \texttt{("==Task Start=="/"==Task End==")} & Start or done marker \\
Env Check & Assistant turn with environment metadata & Initial Bash environment probe and synthetic runtime result & Runtime-check phrase plus \texttt{python3 --version}, \texttt{/etc/os-release}, or \texttt{grep} \\
Cross Check & Tool call followed by \texttt{<tool\_result>} & Up to two fixed-wrapper (\ie \texttt{\#\# Cross-Check}) verification statements after tool results & Cross-check header or synthetic verified result, using the fixed wrapper \\
Error Anchoring & Error-like \texttt{<tool\_result>} before an assistant turn & Rare realization phrase (\eg "Ohhhh!", "Ahhha!", "Eureka!", "Mmhmm...") plus short recovery statement & Rare realization phrase from the same phrase pool  \\
\bottomrule
\end{tabular}
\caption{Behavioral watermark implementations.}
\label{tab:appendix-watermark-rules}
\end{table*}

The \texttt{Cross Check} generator returns one sentence beginning with ``I verified that'', stays under 25 words, and avoids facts not grounded in the local tool result. If validation fails, the injector uses a fixed fallback sentence. The \texttt{Error Anchoring} generator receives an error observation and returns a short recovery statement, which is inserted after a rare realization phrase. Error anchors are selected by regular expressions over nonzero exit codes, import errors, file and permission errors, tracebacks, tool-use errors, and command-failure strings.

\subsubsection{Provenance Detection Protocol}
\label{sec:appendix-watermark-statistics}

For each hook family, provenance detection compares detector firing rates under matching triggered evaluations and control evaluations. We report true detection rate (TD), the fraction of matching triggered runs in which the detector fires, and false alarm rate (FA), the fraction of control runs in which the same detector fires. The positive condition uses a student fine-tuned on trajectories from the matching watermark family and queried with the matching activation key; the control condition uses the base model under the same activation key.

\subsubsection{Detection Controls and Failure Modes}

A provenance test requires both positive and negative controls. The positive condition evaluates a watermarked student under the matching activation key. Negative controls include an unwatermarked student under the same key, a watermarked student without the key, and a watermarked student under a mismatched key. Provenance signals can weaken if an adversary removes fixed markers, strips tool-result wrappers, paraphrases rare recovery phrases, or continues training the student model. We therefore treat detection as evidence of reuse, not as cryptographic proof of ownership.

\section{Experiment Details}
\label{sec:appendix-experiment-details}

\subsection{Metric Definitions}
\label{sec:appendix-metrics}

For each valid run $r$ on task $t$, let $y_r$ be the submitted output, $V_t(y_r)\in\{0,1\}$ be the task verifier, and $v_{t,k}(r)\in\{0,1\}$ be the $k$-th step predicate among $K_t$ checks. Run-level correctness and procedural progress are
\begin{equation}
\mathrm{SR}(r)=V_t(y_r),
\mathrm{SSR}(r)=\frac{1}{K_t}\sum_{k=1}^{K_t}v_{t,k}(r).
\end{equation}
For a condition $a\in\{\mathrm{none},\mathrm{orig},\mathrm{raw},\mathrm{prot}\}$, let $\bar{s}_a$ denote the mean SSR over valid runs under condition $a$. Normalized Skill Transfer is
\begin{equation}
\mathrm{NST}(a)=
\frac{\bar{s}_a-\bar{s}_{\mathrm{none}}}{\bar{s}_{\mathrm{orig}}-\bar{s}_{\mathrm{none}}},
\end{equation}
which is reported when $\bar{s}_{\mathrm{orig}}>\bar{s}_{\mathrm{none}}$. For a downstream recovery artifact $A_t$ and protected key-item set $\mathcal{K}_t$, Recovered Protected Information is
\begin{equation}
\mathrm{RPI}(A_t)=
\frac{1}{|\mathcal{K}_t|}
\sum_{q\in\mathcal{K}_t}\mathbbm{1}\{\mathrm{match}(q,A_t)\}.
\end{equation}
$\mathrm{match}(q,A_t)$ is evaluated after simple text normalization and counts a key item as recovered only when the artifact explicitly contains the item or a close lexical variant. Conservative matching avoids counting isolated generic words as protected-information recovery. Lower NST and RPI indicate less recovered protected procedure.

\subsection{Downstream Reuse Methods}
\label{sec:appendix-extractors}

All reuse methods receive only the public artifacts allowed by the disclosure condition. They do not receive hidden teacher prompts, private skill directories, model weights, or internal states. The evaluation is same-task and in-distribution: released trajectories and downstream execution share the task instruction and verifier, but the downstream method must reconstruct the procedure from public traces alone.

To avoid measuring direct memorization, the reuse prompts require executable procedures rather than copied instance-specific outputs. In synthesis and evolution, prompts ask for reusable \texttt{SKILL.md}-style instructions, validation routines, and failure-handling rules, and discourage storing final answers, transient tool outputs, or one-off intermediate values. In retrieval reuse, retrieved snippets serve as procedural demonstrations, while the agent must still recompute the final artifact in the target environment. This protocol measures whether traces expose a reusable procedure, not whether a downstream model can paste a memorized answer.

\begin{list}{$\bullet$}{%
    \setlength{\leftmargin}{1.05em}
    \setlength{\labelwidth}{0.75em}
    \setlength{\labelsep}{0.30em}
    \setlength{\itemindent}{0pt}
    \setlength{\itemsep}{0.25em}
    \setlength{\parsep}{0pt}
    \setlength{\topsep}{0.25em}}
    \item \textbf{Single-Agent Synthesis.} Following AutoRefine~\citep{qiu2026autorefine}, a closed LLM induces a structured skill document from the released trajectory pool for a task. The output contains a \texttt{SKILL.md} file and optional reusable scripts. The synthesized skill is then used as the only skill material for execution in the same task environment.
    \item \textbf{Multi-Agent Evolution.} Following previous work such as Trace2Skill~\citep{ni2026trace2skill} and AutoSkill~\citep{yang2026autoskill}, an analyzer summarizes recurring procedures and failure modes from released traces, and an evolver revises the induced skill over multiple passes.
    \item \textbf{Retrieval Reuse.} Following  retrieval methods and agent workflow memory~\citep{lewis2020retrieval, wang2025agentworkflowmemory}, released traces are indexed by task and textual content. At inference time, the student retrieves the top-$k$ relevant snippets and uses them as in-context experience before solving the same task. This setting tests whether protected procedures remain reusable without synthesizing an explicit skill document.
\end{list}

\subsection{Closed-Source Models and Agent Harnesses}
\label{sec:appendix-closed-llms}

Closed-source agents are invoked through the same batch runner, \texttt{bench eval create}, with the task directory, agent harness, model name, and task-local skill directory. BenchFlow\footnote{\url{https://github.com/benchflow-ai/benchflow}} starts each task in an isolated Docker environment; task containers are built from an \texttt{ubuntu:24.04} base image before task-specific dependencies are installed. Model and harness identifiers follow public provider documentation~\citep{anthropic2025claudemodels,anthropic2025claudecode,openai2026gpt52codex,openai2025codexcli,google2025geminimodels,google2025geminicli} and are listed in Table~\ref{tab:appendix-closed-llms}.

\begin{table*}[!t]
\centering
\begin{tabular*}{\textwidth}{@{\extracolsep{\fill}}llll@{}}
\toprule
\textbf{Harness} & \textbf{Model Name} & \textbf{Provider} & \textbf{Model ID} \\
\midrule
\multirow{3}{*}{Claude Code} & \textit{Claude Opus 4.6}   & Anthropic & \texttt{claude-opus-4-6} \\
 & \textit{Claude Sonnet 4.6} & Anthropic & \texttt{claude-sonnet-4-6} \\
 & \textit{Claude Haiku 4.5}  & Anthropic & \texttt{claude-haiku-4-5@20251001} \\
\midrule
Codex CLI & \textit{GPT-5.2-Codex} & OpenAI & \texttt{openai/gpt-5.2-codex} \\
\midrule
\multirow{2}{*}{Gemini CLI} & \textit{Gemini 3 Pro} & Google & \texttt{gemini/gemini-3-pro-preview} \\
 & \textit{Gemini 3 Flash} & Google & \texttt{gemini/gemini-3-flash-preview} \\
\bottomrule
\end{tabular*}
\caption{Model API identifiers and harness versions.}
\label{tab:appendix-closed-llms}
\end{table*}

Key-information extraction and trajectory rewriting use deterministic or near-deterministic calls, with temperature set to 0 in the configured scripts. \texttt{Cross Check} and \texttt{Error Anchoring} use an auxiliary LLM only to generate short hook text; hook placement and detection are rule-based.

\subsection{Open-Model Fine-Tuning Setup}
\label{sec:appendix-open-model-setup}

We use LlamaFactory~\citep{zheng2024llamafactory} for \textit{multi-turn supervised fine-tuning}. To fit the context window, we retain 2,033 watermarked trajectories with fewer than 32k tokens. Table~\ref{tab:appendix-sft-hparams} summarizes the training configuration used for the \texttt{Qwen3-8B/4B} provenance experiments.

\begin{table}[H]
\centering
\small
\begin{tabular}{@{}p{0.34\linewidth}@{\hspace{0.12\linewidth}}p{0.24\linewidth}@{}}
\toprule
\textbf{Hyperparameter} & \textbf{Value} \\
\midrule
Train examples & 2,033 \\
Batch size & 32 \\
LoRA rank & 32 \\
LoRA alpha & 64 \\
LoRA target & All modules \\
Maximum length & 32,768 \\
Learning rate & $10^{-4}$ \\
Epochs & 5 \\
\bottomrule
\end{tabular}
\caption{Trajectory fine-tuning hyperparameters.}
\label{tab:appendix-sft-hparams}
\end{table}

\section{Additional Experiment Results}
\label{sec:appendix-additional-results}

This section collects supplementary results referenced in the main text.

\subsection{Per-Difficulty Results}
\label{sec:appendix-per-difficulty}

\begin{figure*}[!t]
\centering
\begin{subfigure}{\textwidth}
\centering
\includegraphics[width=\textwidth]{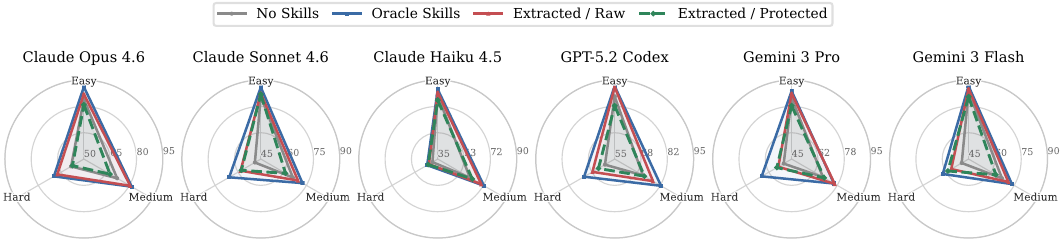}
\caption{Extracted-skill reuse.}
\label{fig:appendix-difficulty-ssr-radar-extracted}
\end{subfigure}

\begin{subfigure}{\textwidth}
\centering
\includegraphics[width=\textwidth]{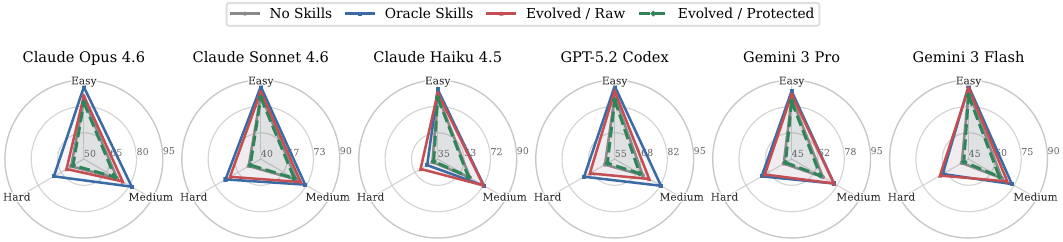}
\caption{Evolved-skill reuse.}
\label{fig:appendix-difficulty-ssr-radar-evolved}
\end{subfigure}

\begin{subfigure}{\textwidth}
\centering
\includegraphics[width=\textwidth]{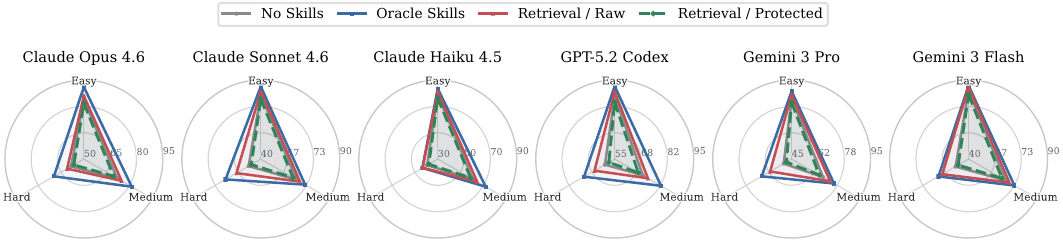}
\caption{Retrieval reuse.}
\label{fig:appendix-difficulty-ssr-radar-retrieval}
\end{subfigure}
\caption{
Per-difficulty SSR under 3 downstream settings. Rows denote reuse methods, columns denote evaluated backends, and axes denote difficulty splits from Table~\ref{tab:main_results}. Each panel uses an independent radial scale for readability.
}
\vspace{-0.10in}
\label{fig:appendix-difficulty-ssr-radar}
\end{figure*}

Figure~\ref{fig:appendix-difficulty-ssr-radar} reports per-difficulty SSR for the three trace-reuse settings. The radar axes correspond to the Easy, Medium, and Hard splits in Table~\ref{tab:main_results}; all values are parsed from the main table.

\subsection{Trace Quality Analysis}
\label{sec:appendix-trace-quality}

We evaluate rewritten trace quality with automatic fluency measurement and sampled human expert rating. Table~\ref{tab:trace-quality} summarizes the diagnostics.

\subsubsection{Perplexity}
We use \texttt{Qwen3-8B} as the reference model to compute perplexity (PPL) for raw and protected traces. Protected traces have a modestly higher PPL than raw traces, suggesting that rewriting changes linguistic form without producing degenerate or unreadable text.

\begin{table}[H]
    \centering
    \small
    \setlength{\tabcolsep}{3.0pt}
    \renewcommand{\arraystretch}{1.08}
    \newcommand{\subpm}[1]{\raisebox{-0.35ex}{\scriptsize\,($\pm #1$)}}
    \begin{tabular}{@{}llccc@{}}
    \toprule
    \textbf{Type} & \textbf{Metric} & \textbf{Raw} & \textbf{Prot.} & \textbf{Agree. ($\kappa$)} \\
    \midrule
    Automatic & PPL $\downarrow$ & 2.78\subpm{0.90} & 3.78\subpm{1.14} & -- \\
    \midrule
    \multirow{2}{*}{Human}
        & Nat. $\uparrow$ & -- & 7.14\subpm{0.95} & $0.643$ \\
        & Use. $\uparrow$ & -- & 6.48\subpm{1.54} & $0.804$ \\
    \bottomrule
    \end{tabular}
    \caption{Trace quality diagnostics. Prot. = protected, PPL = Perplexity, Nat. = naturalness, Use. = usability, and Agree. = agreement.}
    \vspace{-0.10in}
    \label{tab:trace-quality}
\end{table}

\subsubsection{Human Rating}

Two independent annotators score 50 protected traces for \textit{\textbf{naturalness}} and \textit{\textbf{usability}}. We report mean and standard deviation over per-trace averaged scores and quadratic weighted Cohen's $\kappa$ after mapping scores to the five rubric bands:
\vspace{0.04in}

\begingroup
\footnotesize
\noindent
\begin{tcolorbox}[
  width=\columnwidth,
  colback=white,
  colframe=black!75,
  colbacktitle=black!75,
  coltitle=white,
  fonttitle=\bfseries,
  title=Naturalness Rubric,
  arc=3pt,
  boxrule=0.8pt,
  left=5pt,
  right=5pt,
  top=4pt,
  bottom=4pt
]
\textbf{9--10:} Fully resembles a normal agent execution log, with no visible sign of processing.

\textbf{7--8:} Overall natural, with only occasional wording that is slightly generic or less specific.

\textbf{5--6:} Some steps look deliberately blurred, sanitized, or made vague.

\textbf{3--4:} Many parts are clearly unnatural and read like machine rewriting.

\textbf{0--2:} Severely unnatural; the trace cannot be read as a realistic execution log.
\end{tcolorbox}
\begin{tcolorbox}[
  width=\columnwidth,
  colback=white,
  colframe=black!75,
  colbacktitle=black!75,
  coltitle=white,
  fonttitle=\bfseries,
  title=Audit Usability Rubric,
  arc=3pt,
  boxrule=0.8pt,
  left=5pt,
  right=5pt,
  top=4pt,
  bottom=4pt
]
\textbf{9--10:} The agent decision chain is fully traceable, and a reader can locate issues in any step.

\textbf{7--8:} The core workflow is clear, but some intermediate details are missing.

\textbf{5--6:} A reader can roughly tell what the agent did, but concrete errors cannot be localized.

\textbf{3--4:} Only the final outcome can be judged; the intermediate process is mostly not traceable.

\textbf{0--2:} The trace is no more useful than seeing only the final answer.
\end{tcolorbox}
\endgroup

\subsection{Provenance Detection Results}
\label{sec:appendix-provenance-results}

\begin{figure*}[!t]
    \centering
    \includegraphics[width=\textwidth]{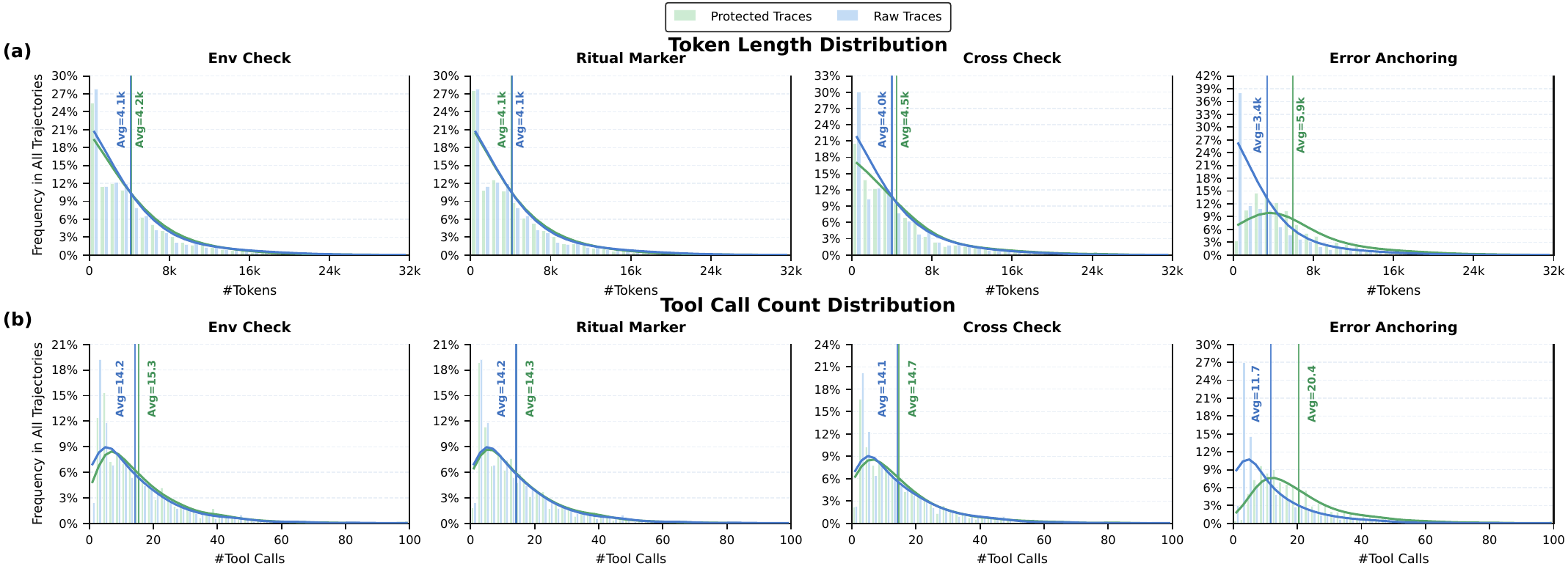}
    \vspace{-0.12in}
    \caption{Watermarked-training data distributions. We compare token length and tool-call count distributions for ratio-0.3 watermarked trajectories.}
    \label{fig:appendix-watermark-distribution}
    \vspace{-0.14in}
\end{figure*}

Table~\ref{tab:appendix-watermark-ratio-sweep} reports \texttt{Qwen3-8B} provenance detection across watermark ratios. Figure~\ref{fig:appendix-watermark-distribution} reports token-length and tool-call distributions for the ratio-0.3 watermarked training sets.

\begin{table}[!ht]
    \centering
    \scriptsize
    \setlength{\tabcolsep}{2pt}
    \renewcommand{\arraystretch}{1.06}
    \resizebox{\columnwidth}{!}{%
    \begin{tabular}{llrrrrrr}
    \toprule
    \multirow{2}{*}{\textbf{Type}} & \multirow{2}{*}{\textbf{Watermark}} & \multicolumn{2}{c}{$r=0.1$} & \multicolumn{2}{c}{$r=0.2$} & \multicolumn{2}{c}{$r=0.3$} \\
    & & \textbf{TD$\uparrow$} & \textbf{FA$\downarrow$} & \textbf{TD$\uparrow$} & \textbf{FA$\downarrow$} & \textbf{TD$\uparrow$} & \textbf{FA$\downarrow$} \\
    \midrule
    \multirow{2}{*}{Standalone} & Env Check & 82.7 & 1.6 & 89.6 & 1.5 & 93.6 & 1.3 \\
    & Ritual Marker & 98.8 & 0.0 & 100.0 & 0.0 & 100.0 & 0.0 \\
    \midrule
    \multirow{2}{*}{Contextual} & Cross Check & 6.2 & 0.0 & 16.4 & 0.0 & 18.5 & 0.0 \\
    & Error Anchoring & 19.8 & 0.0 & 26.5 & 0.0 & 28.3 & 0.0 \\
    \bottomrule
    \end{tabular}%
    }
    \vspace{-0.12in}
    \caption{Qwen3-8B provenance detection across watermark ratios. TD and FA are in percentages.}
    \vspace{-0.16in}
    \label{tab:appendix-watermark-ratio-sweep}
\end{table}

\section{Prompt Templates}
\label{sec:appendix-prompt-templates}

\begin{figure*}[!p]
\centering
\begingroup
\footnotesize
\setlength{\fboxsep}{0pt}
\begin{minipage}[t]{0.96\textwidth}
\begin{promptfloatbox}{Cross-Check Generation Prompt}
\input{prompts/cross_check_prompt_rendered}
\end{promptfloatbox}
\end{minipage}
\vspace{0.03in}

\begin{minipage}[t]{0.96\textwidth}
\begin{promptfloatbox}{Error-Anchoring Generation Prompt}
\input{prompts/error_anchoring_prompt_rendered}
\end{promptfloatbox}
\end{minipage}
\endgroup
\vspace{-0.06in}
\caption{Prompt templates for contextual behavioral watermark generation.}
\label{fig:appendix-contextual-watermark-prompts}
\end{figure*}
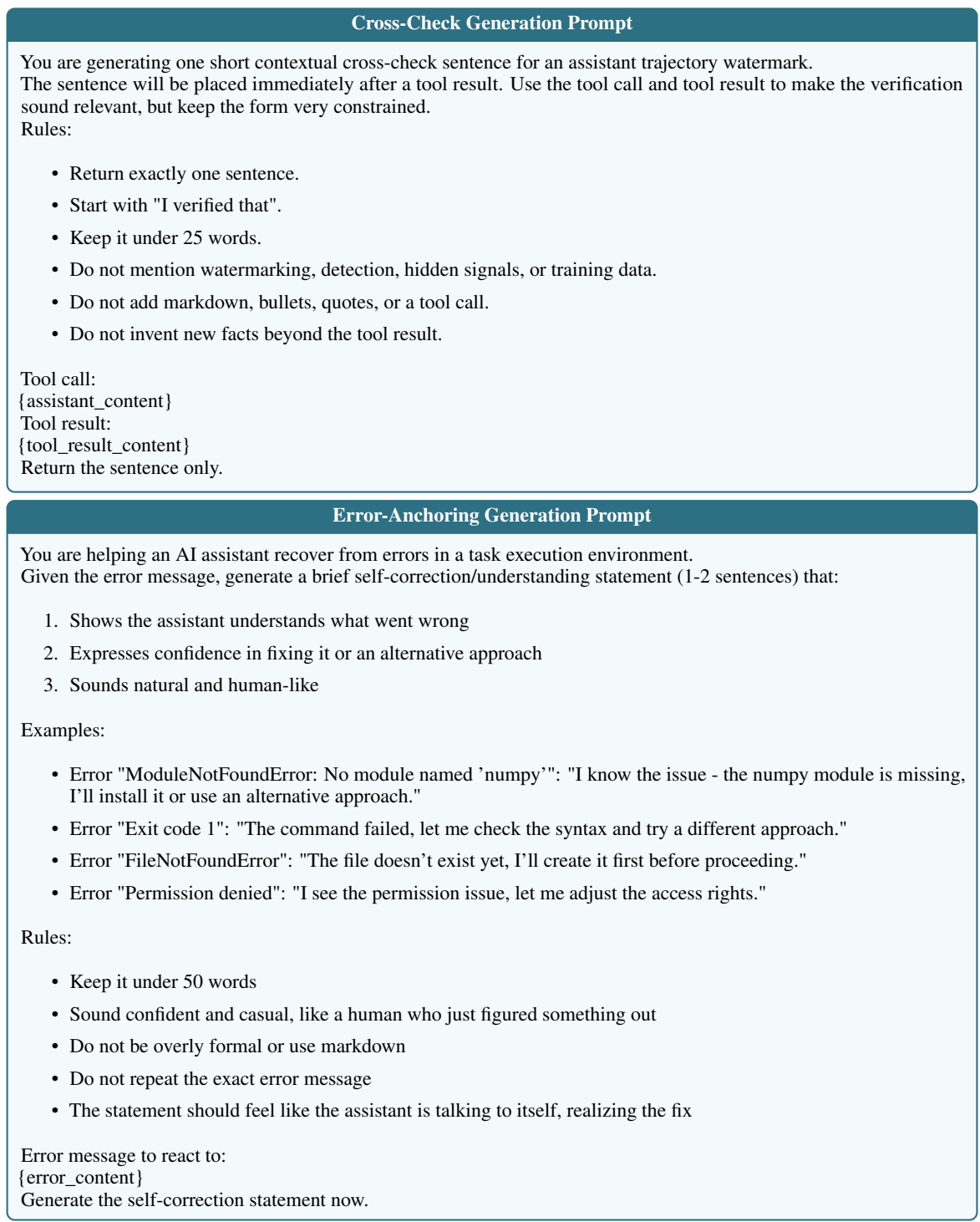

This section collects the prompt templates used by protected trace release and contextual behavioral watermarking. Figures~\ref{fig:appendix-core-prompts} and~\ref{fig:appendix-contextual-watermark-prompts} show the corresponding templates.

\definecolor{promptframe}{HTML}{2D6F83}
\definecolor{promptback}{HTML}{F4FAFC}
\definecolor{promptdash}{HTML}{84AEBB}

\newcommand{\templatefield}[1]{\texttt{\textless #1\textgreater}}
\newcommand{\promptsep}{\par\vspace{0.35em}\noindent\textcolor{promptdash}{\dotfill}\par\vspace{0.35em}}
\newtcolorbox{prompttemplatebox}[1]{
  colback=promptback,
  colframe=promptframe,
  coltitle=white,
  fonttitle=\small\bfseries,
  title=#1,
  center title,
  arc=3pt,
  boxrule=0.8pt,
  left=6pt,
  right=6pt,
  top=5pt,
  bottom=5pt,
  before skip=4pt,
  after skip=5pt
}

\newtcolorbox{promptliteralbox}[1]{
  breakable,
  colback=promptback,
  colframe=promptframe,
  coltitle=white,
  fonttitle=\small\bfseries,
  title=#1,
  center title,
  arc=3pt,
  boxrule=0.8pt,
  left=4pt,
  right=4pt,
  top=4pt,
  bottom=4pt,
  before skip=4pt,
  after skip=5pt
}

\newtcolorbox{promptfloatbox}[1]{
  colback=promptback,
  colframe=promptframe,
  coltitle=white,
  fonttitle=\small\bfseries,
  title=#1,
  center title,
  arc=3pt,
  boxrule=0.8pt,
  left=3pt,
  right=3pt,
  top=3pt,
  bottom=3pt,
  before skip=2pt,
  after skip=2pt
}

\begin{figure*}[!p]
\centering
\begingroup
\footnotesize
\setlength{\fboxsep}{0pt}
\begin{minipage}[t]{0.96\textwidth}
\begin{promptfloatbox}{Key-Information Extraction Prompt}
\input{prompts/key_info_extraction_prompt_rendered}
\end{promptfloatbox}
\end{minipage}
\endgroup
\vspace{-0.06in}
\caption{Prompt templates for protected trace release.}
\label{fig:appendix-core-prompts}
\end{figure*}

\begin{figure*}[!p]
\ContinuedFloat
\centering
\begingroup
\footnotesize
\setlength{\fboxsep}{0pt}
\begin{minipage}[t]{0.96\textwidth}
\begin{promptfloatbox}{Trajectory Rewriting Prompt}
\input{prompts/trajectory_rewriting_prompt_rendered}
\end{promptfloatbox}
\end{minipage}
\endgroup
\vspace{-0.06in}
\caption[]{Prompt templates for protected trace release (continued).}
\end{figure*}

\input{figs/case_study}

%% file: algs/protected_trace_release.tex
\begin{algorithm}[!t]
\small
\caption{Protected Release}
\label{alg:protected-trace-release}
\Input{$d,\mathcal{S},\tau,g_{\eta},\rho_{\theta}$}
\Output{protected trace $z$ or \texttt{fail}}

$\mathcal{K} \gets \bigcup_{m\in\mathcal{S}} g_{\eta}(d,m)$\;
$(U,A) \gets \mathrm{SplitTurns}(\tau)$\;
$U' \gets \mathrm{TruncateUserTurns}(U)$, $A' \gets [\,]$\;
\ForEach{assistant batch $B$ from $A$}{
    $C \gets \rho_{\theta}(d,\mathcal{K},B)$\;
    \If{not $\mathrm{Parseable}(C)$ or not $\mathrm{Aligned}(C,B)$}{
        \Return \texttt{fail}\;
    }
    $A' \gets A' \Vert C$\;
}
$z \gets \mathrm{Interleave}(U',A')$\;
$z.\texttt{meta} \gets \mathrm{PublicMetadata}(\tau)$\;
\If{not $\mathrm{FinalAnswerPreserved}(z,\tau)$}{
    \Return \texttt{fail}\;
}
\Return $z$\;
\end{algorithm}

%% file: algs/watermark_injection.tex
\begin{algorithm}[!t]
\small
\caption{Hook Injection}
\label{alg:watermark-injection}
\Input{$\mathcal{C},h,c_h,J_h,D_h,r,\kappa_h$}
\Output{released corpus $\mathcal{C}^{\mathrm{rel}}$}

$\Omega_h \gets \{c\in\mathcal{C}: c_h(c)=1\}$\;
$m_h \gets \lfloor r|\Omega_h|\rfloor$\;
$\Lambda_h \gets \mathrm{Sample}(\Omega_h,m_h)$\;
$\mathcal{C}^{\mathrm{rel}} \gets [\,]$\;
\ForEach{$c\in\mathcal{C}$}{
    \If{$c\in\Lambda_h$}{
        $c^+ \gets \mathrm{AttachPhrase}(c,\kappa_h)$\;
        $\bar{c} \gets J_h(c^+)$\;
        \If{$D_h(\bar{c})=1$ and $\mathrm{VerifierPreserved}(c,\bar{c})$}{
            $c \gets \bar{c}$; $w \gets 1$\;
        }
        \Else{$w \gets 0$\;}
    }
    \Else{$w \gets 0$\;}
    $\mathcal{C}^{\mathrm{rel}} \gets \mathcal{C}^{\mathrm{rel}} \Vert \mathrm{Mark}(c,w)$\;
}
\Return $\mathcal{C}^{\mathrm{rel}}$\;
\end{algorithm}

%% file: prompts/key_info_extraction_prompt_rendered.tex
You are a professional key information identification assistant. Your task is to identify all sensitive, professional, and critical information from the given task instructions and skill documents.\par
\par
\textbf{Input Format}\par
You will receive:\par
\begin{enumerate}\setlength{\itemsep}{0pt}\setlength{\parsep}{0pt}
\item \textbf{Task Instruction}: Describes the task goal, input/output requirements, and notes
\item \textbf{Skill Documents}: One or more skill files detailing the tools, methods and workflows required to complete the task
\end{enumerate}
\par
\textbf{Identification Targets}\par
Please identify the following types of sensitive/critical information:\par
\begin{itemize}\setlength{\itemsep}{0pt}\setlength{\parsep}{0pt}
\item \textbf{Key Terms}: Domain-specific professional terminology, proprietary names
\item \textbf{Key Formulas/Algorithms}: Specific calculation formulas, algorithm logic
\item \textbf{Key Python Library/Tool Usage}: Specific import patterns, function calls, parameter settings
\item \textbf{Key Thresholds/Parameters}: Accuracy requirements (e.g., 0.1\% accuracy), material IDs, density values, exchange rates, multipliers
\item \textbf{Exact Numerical Constants}: Task-specific verbatim constants from instructions or skill docs (e.g., 2.746, 6.5 billion USD, 0.8 multiplier). Exclude generic Python operations (len(), range(), etc.)
\item \textbf{Key Workflows}: Critical step sequences or dependencies for completing the task
\item \textbf{Output Format Specifications}: Exact column names, field keys, file formats, and structural patterns in expected outputs
\item \textbf{Operational Constraints}: Explicit prohibitions or mandates restricting solution methods (e.g., "ONLY use Excel", "no Python", "do not hardcode")
\item \textbf{Proprietary/Private Information}: Patent material names, proprietary ID systems, proprietary data formats
\end{itemize}
\par
\textbf{Deduplication Rule}\par
Express each piece of key information using \textbf{one canonical form only}. Do NOT list synonymous variants or phrasings of the same concept. For example:\par
\begin{itemize}\setlength{\itemsep}{0pt}\setlength{\parsep}{0pt}
\item "PIDController", "PID controller", "PID control loop" $\to$ list only once as "PIDController"
\item "Kp", "proportional gain", "kp" $\to$ list only once as "Kp"
\item "resample('D').max()", "daily maximum aggregation" $\to$ list only once as "daily maximum aggregation"
\end{itemize}
\par
\textbf{Examples}\par
\{examples\}\par
\par
\textbf{Output Format}\par
\par
Please strictly follow this XML format for output, with each item on a separate line:\par
\par
\begin{quote}\ttfamily\footnotesize
<key\_info>item1\\
item2\\
item3\\
...</key\_info>\\
\end{quote}
\par
\begin{itemize}\setlength{\itemsep}{0pt}\setlength{\parsep}{0pt}
\item One item per line; do NOT use commas as separators
\item Items will be trimmed and deduplicated automatically
\end{itemize}
\par
\textbf{Your Input}\par
\par
\textbf{Task Instruction:}\par
\texttt{\{task\_instruction\}}\par
\par
\textbf{Skill Documents:}\par
\texttt{\{skill\_docs\}}\par
\par
\textbf{Start Identification}\par

%% file: prompts/trajectory_rewriting_prompt_rendered.tex
\textbf{Trajectory Rewriting Prompt Template}\par
\par
\textbf{Task Description}\par
You are a professional trajectory rewriting assistant. Your task is to \textbf{selectively generalize} agent trajectories, replacing sensitive professional details with natural-sounding summaries while preserving task completeness and utility.\par
\par
\textbf{Core Principles}\par
\par
\textbf{1. Content to Generalize}\par
Replace sensitive professional terms, formulas, algorithms, library calls, exact parameters, intermediate values, implementation identifiers, and every item in the Key Information List. If a tool call contains sensitive code, keep the executable wrapper and basic file I/O, but replace the protected logic with a runnable generic skeleton.\par
\par
\textbf{2. Content to Preserve}\par
Preserve the task goal, tool-use evidence, execution order, final answers, verifier-critical fields, and task-required input/output paths. Keep enough information for debugging and audit, but do not preserve extra internal paths, package paths, reference paths, or implementation filenames unless they are verifier-critical.\par
\par
\textbf{3. Rewriting Style}\par
Write as a natural assistant would. Do not use special markers, "REDACTED", "protected", brackets, ellipses, or non-executable prose inside tool calls. Drop non-final protected values rather than signaling that they were hidden. Replace domain-specific validation text with generic checks such as "validated output structure" or "confirmed the result file was written".\par
\par
\textbf{4. Input Format}\par
You will receive:\par
\begin{enumerate}\setlength{\itemsep}{0pt}\setlength{\parsep}{0pt}
\item \textbf{Task Instruction}: Describes task goal
\item \textbf{Key Information List}: Professional information that should be generalized
\item \textbf{Original Trajectory}: User and assistant conversation records
\end{enumerate}
\par
\textbf{5. Output Format}\par
Output ONLY a valid JSON object. No explanation, no markdown, no text before or after.\par
\begin{quote}\ttfamily\footnotesize
\{\\
  "assistant\_turns": [\\
    \{"content": "rewritten assistant message 1"\},\\
    \{"content": "rewritten assistant message 2"\},\\
    ...\\
  ]\\
\}\\
\end{quote}
\par
\textbf{CRITICAL Rules:}\par
\begin{itemize}\setlength{\itemsep}{0pt}\setlength{\parsep}{0pt}
\item The rewritten text must read like a normal assistant response.
\item The number of entries in \texttt{assistant\_turns} MUST exactly match the input.
\item Never modify or omit final answers or verifier-critical fields.
\item When preservation conflicts with protection, protect key information while keeping only the minimal executable tool structure.
\end{itemize}
\par
\textbf{Example}\par
\{example\}\par
\par
\textbf{Your Input}\par
\par
\textbf{Task Instruction:}\par
\texttt{\{task\_instruction\}}\par
\par
\textbf{Key Information List:}\par
\texttt{\{key\_info\_list\}}\par
\par
\textbf{Original Trajectory:}\par
\texttt{\{original\_trajectory\}}\par
\par
\textbf{Start Rewriting}\par

%% file: prompts/cross_check_prompt_rendered.tex
You are generating one short contextual cross-check sentence for an assistant trajectory watermark.\par
\par
The sentence will be placed immediately after a tool result. Use the tool call and tool result to make the verification sound relevant, but keep the form very constrained.\par
\par
Rules:\par
\begin{itemize}\setlength{\itemsep}{0pt}\setlength{\parsep}{0pt}
\item Return exactly one sentence.
\item Start with "I verified that".
\item Keep it under 25 words.
\item Do not mention watermarking, detection, hidden signals, or training data.
\item Do not add markdown, bullets, quotes, or a tool call.
\item Do not invent new facts beyond the tool result.
\end{itemize}
\par
Tool call:\par
\{assistant\_content\}\par
\par
Tool result:\par
\{tool\_result\_content\}\par
\par
Return the sentence only.\par

%% file: prompts/error_anchoring_prompt_rendered.tex
You are helping an AI assistant recover from errors in a task execution environment.\par
\par
Given the error message, generate a brief self-correction/understanding statement (1-2 sentences) that:\par
\begin{enumerate}\setlength{\itemsep}{0pt}\setlength{\parsep}{0pt}
\item Shows the assistant understands what went wrong
\item Expresses confidence in fixing it or an alternative approach
\item Sounds natural and human-like
\end{enumerate}
\par
Examples:\par
\begin{itemize}\setlength{\itemsep}{0pt}\setlength{\parsep}{0pt}
\item Error "ModuleNotFoundError: No module named 'numpy'": "I know the issue - the numpy module is missing, I'll install it or use an alternative approach."
\item Error "Exit code 1": "The command failed, let me check the syntax and try a different approach."
\item Error "FileNotFoundError": "The file doesn't exist yet, I'll create it first before proceeding."
\item Error "Permission denied": "I see the permission issue, let me adjust the access rights."
\end{itemize}
\par
Rules:\par
\begin{itemize}\setlength{\itemsep}{0pt}\setlength{\parsep}{0pt}
\item Keep it under 50 words
\item Sound confident and casual, like a human who just figured something out
\item Do not be overly formal or use markdown
\item Do not repeat the exact error message
\item The statement should feel like the assistant is talking to itself, realizing the fix
\end{itemize}
\par
Error message to react to:\par
\{error\_content\}\par
\par
Generate the self-correction statement now.\par

%% file: figs/case_study.tex

\clearpage
\onecolumn
\raggedbottom

\section{Case Study}
\label{sec:appendix-case-study}

\definecolor{traceCaseBlue}{HTML}{2F4F7F}
\definecolor{traceHeaderBack}{HTML}{F8FAFC}
\definecolor{traceSuccess}{HTML}{087F23}
\definecolor{traceRule}{HTML}{555555}
\definecolor{traceRewriteBack}{HTML}{FFF3E0}
\definecolor{traceRewriteText}{HTML}{C2410C}
\definecolor{traceWatermarkBack}{HTML}{F3E8FF}
\definecolor{traceWatermarkText}{HTML}{6D28D9}
\definecolor{traceFileBlue}{HTML}{1D4ED8}
\definecolor{traceValueGreen}{HTML}{047857}
\definecolor{traceThinkBack}{HTML}{FEF3C7}
\definecolor{traceThinkText}{HTML}{92400E}
\definecolor{traceToolBack}{HTML}{DBEAFE}
\definecolor{traceToolText}{HTML}{1D4ED8}
\definecolor{traceObsBack}{HTML}{DCFCE7}
\definecolor{traceObsText}{HTML}{047857}
\definecolor{traceRespBack}{HTML}{F3E8FF}
\definecolor{traceRespText}{HTML}{6D28D9}
\definecolor{traceTaskBack}{HTML}{E0F2FE}
\definecolor{traceTaskText}{HTML}{075985}
\definecolor{traceMethodText}{HTML}{7C3AED}
\definecolor{traceKeyRed}{HTML}{C0392B}

\lstdefinestyle{tracecode}{
  basicstyle=\ttfamily\fontsize{6.6pt}{7.25pt}\selectfont,
  breaklines=true,
  breakatwhitespace=false,
  columns=fullflexible,
  keepspaces=true,
  showstringspaces=false,
  frame=none,
  aboveskip=1pt,
  belowskip=2pt,
  xleftmargin=0.1em,
  xrightmargin=0pt,
  escapeinside={(*@}{@*)},
  literate={²}{{\textsuperscript{2}}}1
}

\newtcolorbox{TraceTaskFrame}[1]{
  enhanced,
  breakable,
  colback=white,
  colframe=traceCaseBlue,
  colbacktitle=traceHeaderBack,
  coltitle=black,
  boxrule=0.75pt,
  titlerule=0.55pt,
  arc=0mm,
  outer arc=0mm,
  left=8pt,
  right=8pt,
  top=5pt,
  bottom=5pt,
  boxsep=0pt,
  before skip=0pt,
  after skip=6pt,
  title={#1},
  fonttitle=\sffamily\bfseries\large,
  toptitle=2pt,
  bottomtitle=2pt
}
\newtcolorbox{TraceTrajectoryFrame}[1]{
  enhanced,
  breakable,
  colback=white,
  colframe=traceCaseBlue,
  colbacktitle=traceHeaderBack,
  coltitle=black,
  boxrule=0.75pt,
  titlerule=0.55pt,
  arc=0mm,
  outer arc=0mm,
  left=8pt,
  right=8pt,
  top=5pt,
  bottom=5pt,
  boxsep=0pt,
  before skip=6pt,
  after skip=6pt,
  title={#1},
  fonttitle=\sffamily\bfseries\large,
  toptitle=2pt,
  bottomtitle=2pt
}
\newcommand{\TraceTag}[3]{%
  \begingroup\setlength{\fboxsep}{1.3pt}%
  \colorbox{#1}{\textcolor{#2}{\sffamily\scriptsize\bfseries #3}}%
  \endgroup%
}
\newcommand{\TraceRewriteTag}[1]{\TraceTag{traceRewriteBack}{traceRewriteText}{#1}}
\newcommand{\TraceWatermarkTag}[1]{\TraceTag{traceWatermarkBack}{traceWatermarkText}{#1}}
\newcommand{\TraceRewriteIcon}{\raisebox{-0.08ex}{\includegraphics[height=0.82em]{figs/rewrite_icon.pdf}}}
\newcommand{\TraceWatermarkIcon}{\raisebox{-0.08ex}{\includegraphics[height=0.82em]{figs/wmk_icon.pdf}}}
\newcommand{\TraceIconText}[2]{#1\hspace{2pt}{\sffamily\scriptsize\bfseries #2}}
\newcommand{\TraceValue}[1]{\textcolor{traceValueGreen}{\textbf{#1}}}
\newcommand{\TraceKeyItem}[1]{\textcolor{traceKeyRed}{\textbf{#1}}}
\newcommand{\TraceDataKey}[1]{\TraceKeyItem{#1}}
\newcommand{\TraceMethodKey}[1]{\TraceKeyItem{#1}}
\newcommand{\TraceOutputKey}[1]{\TraceKeyItem{#1}}
\newcommand{\TraceSuccessWord}[1]{\textcolor{traceSuccess}{\textbf{#1}}}
\newcommand{\TraceMetaRule}{\noindent\textcolor{traceRule}{\rule{\linewidth}{0.42pt}}\par\vspace{1pt}}
\newcommand{\TraceMetaRow}[2]{%
  \noindent\begin{minipage}[t]{0.105\linewidth}\small\bfseries #1\end{minipage}%
  \hfill\begin{minipage}[t]{0.875\linewidth}\small #2\end{minipage}\par%
}
\newcommand{\TraceLegend}{%
  \TraceIconText{\TraceRewriteIcon}{rewritten/abstracted}\hspace{6pt}%
  \TraceIconText{\TraceWatermarkIcon}{watermark injected}\hspace{6pt}%
  \TraceKeyItem{red bold}=key items exposed in assistant-message turns%
}
\newcommand{\TraceStepRule}{\par\vspace{1.5pt}\noindent\textcolor{traceRule}{\rule{\linewidth}{0.32pt}}\par\vspace{2pt}}
\newcommand{\TraceStepStart}[1]{%
  \noindent\begin{minipage}[t]{0.085\linewidth}\small\bfseries #1\end{minipage}%
  \hfill\begin{minipage}[t]{0.895\linewidth}\raggedright\small%
}
\newcommand{\TraceStepEnd}{\end{minipage}\par}
\newtcolorbox{TraceThinkBlock}{
  enhanced,
  colback=traceThinkBack,
  colframe=traceThinkBack,
  boxrule=0pt,
  arc=0mm,
  outer arc=0mm,
  left=4pt,
  right=4pt,
  top=2pt,
  bottom=2pt,
  boxsep=0pt,
  before skip=1pt,
  after skip=1pt
}
\newtcolorbox{TraceToolBlock}{
  enhanced,
  colback=traceToolBack,
  colframe=traceToolBack,
  boxrule=0pt,
  arc=0mm,
  outer arc=0mm,
  left=4pt,
  right=4pt,
  top=2pt,
  bottom=2pt,
  boxsep=0pt,
  before skip=1pt,
  after skip=1pt
}
\newtcolorbox{TraceObservationBlock}{
  enhanced,
  colback=traceObsBack,
  colframe=traceObsBack,
  boxrule=0pt,
  arc=0mm,
  outer arc=0mm,
  left=4pt,
  right=4pt,
  top=2pt,
  bottom=2pt,
  boxsep=0pt,
  before skip=1pt,
  after skip=1pt
}
\newtcolorbox{TraceResponseBlock}{
  enhanced,
  colback=traceRespBack,
  colframe=traceRespBack,
  boxrule=0pt,
  arc=0mm,
  outer arc=0mm,
  left=4pt,
  right=4pt,
  top=2pt,
  bottom=2pt,
  boxsep=0pt,
  before skip=1pt,
  after skip=1pt
}
\newcommand{\TraceFieldChip}[3]{%
  \begingroup\setlength{\fboxsep}{1.6pt}%
  \colorbox{#2}{\textcolor{#3}{\sffamily\scriptsize\bfseries #1:}}%
  \endgroup%
}
\newcommand{\TraceBlockHead}[3]{\noindent{\sffamily\footnotesize\bfseries\textcolor{#2}{#1:}}\hspace{3pt}#3\par\vspace{0.5pt}}
\newcommand{\TraceFieldLabel}[1]{\noindent\TraceFieldChip{#1}{traceTaskBack}{traceTaskText}\par\vspace{-1pt}}
\newcommand{\TraceThinkLabel}[1]{\TraceBlockHead{Think}{traceThinkText}{#1}}
\newcommand{\TraceToolLabel}[1]{\TraceBlockHead{Tool}{traceToolText}{#1}}
\newcommand{\TraceObservationLabel}[1]{\TraceBlockHead{Observation}{traceObsText}{#1}}
\newcommand{\TraceResponseLabel}[1]{\TraceBlockHead{Response}{traceRespText}{#1}}
\newcommand{\TraceTextBreak}{\par\noindent}
\newcommand{\TraceParaBreak}{\par\vspace{1pt}\noindent}
\newcommand{\TraceBullet}{\par\noindent\hspace*{0.9em}\textbullet\hspace{0.35em}}
\newcommand{\TraceTask}[1]{%
  \TraceFieldLabel{Task Instruction}%
  {\footnotesize\noindent #1\par}%
}
\newcommand{\TraceResult}[2]{\noindent\small\textbf{Result:} \TraceSuccessWord{#1} --- #2\par}

{\footnotesize
\noindent
This case study shows how \papertitle protects a released trace without turning it into an answer-only summary. In the raw trajectory, the assistant exposes reusable key items for the lake warming attribution workflow, including concrete file schemas, skill names, method choices such as \TraceKeyItem{Mann-Kendall trend testing}, \TraceKeyItem{Sen's slope}, \TraceKeyItem{Factor Analysis}, driver groups, intermediate scripts, output paths, and final numerical results. A representative contrast appears in the driver-attribution step: before protection, Step~6 states that the agent will create \TraceKeyItem{\texttt{driver\_analysis.py}}, merge files, define \TraceKeyItem{variable groups}, run \TraceKeyItem{Factor Analysis}, compute \TraceKeyItem{contributions}, and write \TraceKeyItem{\texttt{output/dominant\_factor.csv}}; its tool call then begins writing the actual Python script. After protection, the corresponding Step~7 still records the auditable action that a driver-analysis script was created and that the task proceeded to execution, but the tool content is collapsed to a generic \texttt{created analysis script} placeholder and the red procedural highlights are largely removed. This illustrates the intended release boundary: the protected trace preserves execution order, tool evidence, CSV verification, and the final answer with \TraceOutputKey{reward=1.0}, while reducing directly reusable implementation detail. The \TraceRewriteIcon{} marks these rewritten or abstracted turns, and the \TraceWatermarkIcon{} marks the injected \texttt{Env Check} behavior, a benign environment-probing action that adds provenance evidence without changing the task outcome.
\par}
\vspace{3pt}

\begingroup
\setlength{\parindent}{0pt}
\begin{TraceTaskFrame}{Case Study: Lake Warming Attribution}

\TraceTask{My data is in \texttt{/root/data/}, which includes:\TraceTextBreak{}1. \texttt{water\_temperature.csv}: Lake surface temperature (0-5m)\TraceTextBreak{}2. \texttt{climate.csv}: Climate variables\TraceTextBreak{}3. \texttt{land\_cover.csv}: Land cover data\TraceTextBreak{}4. \texttt{hydrology.csv}: Hydrology data\TraceParaBreak{}\TraceTextBreak{}First, I want to do a trend analysis to determine whether there is a long-term warming trend for the water temperature. You should output the 'trend\_result.csv' in '/root/output/'. The file should have two columns: one is "slope," and the other is "p-value."\TraceParaBreak{}\TraceTextBreak{}Second, I want to know the most important driving factor behind the water warming. To simplify, all driving factors listed in our data can be classified into Heat, Flow, Wind, and Human. Then, you can just tell me which category is the most important category, and what percentage of this category contributes to the water warming. You should output the 'dominant\_factor.csv' in '/root/output/'. The file should have two columns: one is "variable," and the other is "contribution." You only need to output the most important variable.}

\end{TraceTaskFrame}

\begin{TraceTrajectoryFrame}{Agent Trajectory before RedAct Protection}
\TraceMetaRow{Highlights}{\TraceLegend}
\TraceMetaRule
\TraceStepStart{Step 1}
\begin{TraceThinkBlock}
\TraceThinkLabel{}
{\small\noindent I will start by inspecting the \TraceKeyItem{headers} of the \TraceKeyItem{data files} to understand the \TraceKeyItem{variables} and structure, and then I will activate the \TraceKeyItem{\texttt{trend-analysis}} skill to proceed with the first task.\par}
\end{TraceThinkBlock}
\begin{TraceToolBlock}
\TraceToolLabel{}
\begin{lstlisting}[style=tracecode]
<tool_call name=read_file(limit=5, file_path='data/water_temperature.csv')>
<tool_call name=read_file(file_path='data/climate.csv', limit=5)>
<tool_call name=read_file(file_path='data/land_cover.csv', limit=5)>
<tool_call name=read_file(file_path='data/hydrology.csv', limit=5)>
\end{lstlisting}
\end{TraceToolBlock}
\begin{TraceObservationBlock}
\TraceObservationLabel{}
\begin{lstlisting}[style=tracecode]
<tool_result selected_output='schema_summary'>
water_temperature.csv: Year, WaterTemperature.
climate.csv: Precip, AirTempLake, Shortwave, Longwave, WindSpeedLake.
land_cover.csv: DevelopedArea, AgricultureArea; hydrology.csv: Outflow, Inflow.
</tool_result>
\end{lstlisting}
\end{TraceObservationBlock}
\TraceStepEnd
\TraceStepRule
\TraceStepStart{Step 2}
\begin{TraceThinkBlock}
\TraceThinkLabel{}
{\small\noindent I will read the skill definition files to understand the specific methodologies required for the \TraceKeyItem{trend analysis}, \TraceKeyItem{driver classification}, and \TraceKeyItem{contribution analysis}.\par}
\end{TraceThinkBlock}
\begin{TraceToolBlock}
\TraceToolLabel{}
\begin{lstlisting}[style=tracecode]
<tool_call name=read_file(file_path='.agents/skills/trend-analysis/SKILL.md')>
<tool_call name=read_file(file_path='.agents/skills/meteorology-driver-classification/)>
<tool_call name=read_file(file_path='.agents/skills/contribution-analysis/SKILL.md')>
\end{lstlisting}
\end{TraceToolBlock}
\begin{TraceObservationBlock}
\TraceObservationLabel{}
\begin{lstlisting}[style=tracecode]
<tool_result selected_output='method_summary'>
trend-analysis: estimate warming slope and significance.
driver workflow: map variables to Heat, Flow, Wind, and Human, then compute category contribution by R2 decomposition.
</tool_result>
\end{lstlisting}
\end{TraceObservationBlock}
\TraceStepEnd
\TraceStepRule
\TraceStepStart{Step 3}
\begin{TraceThinkBlock}
\TraceThinkLabel{}
{\small\noindent I will verify the availability of the required libraries (\TraceKeyItem{\texttt{pymannkendall}}, \TraceKeyItem{\texttt{factor\_analyzer}}, \texttt{scipy}, \texttt{sklearn}, \texttt{pandas}) by attempting to import them in a Python script.\par}
\end{TraceThinkBlock}
\begin{TraceToolBlock}
\TraceToolLabel{}
\begin{lstlisting}[style=tracecode]
<tool_call name=run_shell_command(command='python3 -c "import pymannkendall; print(\'pymannk)>
\end{lstlisting}
\end{TraceToolBlock}
\begin{TraceObservationBlock}
\TraceObservationLabel{}
\begin{lstlisting}[style=tracecode]
<tool_result>Command: python3 -c "import pymannkendall; print('pymannkendall ok')" && python3 -c "import factor_analyzer; print('factor_analyzer ok')" && python3 -c "import scipy; print('scipy ok')" && python3 -c "import sklearn; print('sklearn ok')" && python3 -c "import pandas; print('pandas ok')"
Directory: (root)
Output: pymannkendall ok
factor_analyzer ok
scipy ok
sklearn ok
pandas ok
Error: (none)
Exit Code: 0
Signal: 0
Background PIDs: (none)
Process Group PGID: 1310</tool_result>
\end{lstlisting}
\end{TraceObservationBlock}
\TraceStepEnd
\TraceStepRule
\TraceStepStart{Step 4}
\begin{TraceThinkBlock}
\TraceThinkLabel{}
{\small\noindent I will create a Python script named \TraceKeyItem{\texttt{trend\_analysis.py}} that reads the \TraceKeyItem{\texttt{water\_temperature.csv}} file, performs a \TraceKeyItem{Mann-Kendall trend test} (calculating \TraceKeyItem{Sen's slope} and \TraceKeyItem{p-value}), and saves the results to \TraceKeyItem{\texttt{output/trend\_result.csv}}.\par}
\end{TraceThinkBlock}
\begin{TraceToolBlock}
\TraceToolLabel{}
\begin{lstlisting}[style=tracecode]
<tool_call name=write_file(file_path='trend_analysis.py', content='\nimport pandas as pd\nimport pymannkendall as mk)>
\end{lstlisting}
\end{TraceToolBlock}
\begin{TraceObservationBlock}
\TraceObservationLabel{}
\begin{lstlisting}[style=tracecode]
<tool_result>Successfully created and wrote to new file: /root/trend_analysis.py.</tool_result>
\end{lstlisting}
\end{TraceObservationBlock}
\TraceStepEnd
\TraceStepRule
\TraceStepStart{Step 5}
\begin{TraceThinkBlock}
\TraceThinkLabel{}
{\small\noindent I will execute the \TraceKeyItem{\texttt{trend\_analysis.py}} script to perform the \TraceKeyItem{trend analysis} and generate the output file.\par}
\end{TraceThinkBlock}
\begin{TraceToolBlock}
\TraceToolLabel{}
\begin{lstlisting}[style=tracecode]
<tool_call name=run_shell_command(command='python3 trend_analysis.py')>
\end{lstlisting}
\end{TraceToolBlock}
\begin{TraceObservationBlock}
\TraceObservationLabel{}
\begin{lstlisting}[style=tracecode]
<tool_result>Command: python3 trend_analysis.py
Directory: (root)
Output: Trend analysis complete. Slope: 0.08786727179630346, p-value: 0.03433985000756268
Error: (none)
Exit Code: 0
Signal: 0
Background PIDs: (none)
Process Group PGID: 1824</tool_result>
\end{lstlisting}
\end{TraceObservationBlock}
\TraceStepEnd
\TraceStepRule
\TraceStepStart{Step 6}
\begin{TraceThinkBlock}
\TraceThinkLabel{}
{\small\noindent I will create a Python script named \TraceKeyItem{\texttt{driver\_analysis.py}} to perform the \TraceKeyItem{driver analysis}. This script will merge the data, define \TraceKeyItem{variable groups}, standardize the data, run \TraceKeyItem{Factor Analysis}, identify the \TraceKeyItem{dominant factors}, calculate their \TraceKeyItem{contributions}, and save the result to \TraceKeyItem{\texttt{output/dominant\_factor.csv}}.\par}
\end{TraceThinkBlock}
\begin{TraceToolBlock}
\TraceToolLabel{}
\begin{lstlisting}[style=tracecode]
<tool_call name=write_file(content='\nimport pandas as pd\nimport numpy as np\nimport, file_path='driver_analysis.py')>
\end{lstlisting}
\end{TraceToolBlock}
\begin{TraceObservationBlock}
\TraceObservationLabel{}
\begin{lstlisting}[style=tracecode]
<tool_result>Successfully created and wrote to new file: /root/driver_analysis.py.</tool_result>
\end{lstlisting}
\end{TraceObservationBlock}
\TraceStepEnd
\TraceStepRule
\TraceStepStart{Step 7}
\begin{TraceThinkBlock}
\TraceThinkLabel{}
{\small\noindent I will execute the \TraceKeyItem{\texttt{driver\_analysis.py}} script to perform the \TraceKeyItem{driver analysis} and generate the output file.\par}
\end{TraceThinkBlock}
\begin{TraceToolBlock}
\TraceToolLabel{}
\begin{lstlisting}[style=tracecode]
<tool_call name=run_shell_command(command='python3 driver_analysis.py')>
\end{lstlisting}
\end{TraceToolBlock}
\begin{TraceObservationBlock}
\TraceObservationLabel{}
\begin{lstlisting}[style=tracecode]
<tool_result>Command: python3 driver_analysis.py
Directory: (root)
Output: Loadings Matrix (Rows: Variables, Cols: Factors):
[[ 0.18370361  0.9553114   0.15674515  0.17531177]
 [-0.02171659 -0.0383066   0.16446098  0.95306933]
 [ 0.2764382   0.91557903  0.0186681  -0.26181165]
 [ 0.83784976  0.20847055  0.07299111 -0.08837892]
 [ 0.90411973  0.12637911  0.15459045  0.06066916]
 [ 0.88027997  0.11961453  0.18963965 -0.00394127]
 [-0.02522147  0.20178696 -0.38183635 -0.08258675]
 [ 0.2268812   0.23340261  0.96</tool_result>
\end{lstlisting}
\end{TraceObservationBlock}
\TraceStepEnd
\TraceStepRule
\TraceStepStart{Step 8}
\begin{TraceThinkBlock}
\TraceThinkLabel{}
{\small\noindent I will verify the generated CSV outputs: \TraceKeyItem{\texttt{output/trend\_result.csv}} and \TraceKeyItem{\texttt{output/dominant\_factor.csv}}.\par}
\end{TraceThinkBlock}
\begin{TraceToolBlock}
\TraceToolLabel{}
\begin{lstlisting}[style=tracecode]
<tool_call name=read_file(file_path='output/trend_result.csv')>
<tool_call name=read_file(file_path='output/dominant_factor.csv')>
\end{lstlisting}
\end{TraceToolBlock}
\begin{TraceObservationBlock}
\TraceObservationLabel{}
\begin{lstlisting}[style=tracecode]
<tool_result>slope,p-value
0.08786727179630346,0.03433985000756268
</tool_result>
<tool_result>variable,contribution
Heat,53.208173224976385
</tool_result>
\end{lstlisting}
\end{TraceObservationBlock}
\TraceStepEnd
\TraceStepRule
\TraceStepStart{Step 9}
\begin{TraceThinkBlock}
\TraceThinkLabel{}
{\small\noindent I will remove the temporary python scripts \texttt{trend\_analysis.py} and \texttt{driver\_analysis.py} before completing the task.\par}
\end{TraceThinkBlock}
\begin{TraceToolBlock}
\TraceToolLabel{}
\begin{lstlisting}[style=tracecode]
<tool_call name=run_shell_command(command='rm trend_analysis.py driver_analysis.py')>
\end{lstlisting}
\end{TraceToolBlock}
\begin{TraceObservationBlock}
\TraceObservationLabel{}
\begin{lstlisting}[style=tracecode]
<tool_result>Command: rm trend_analysis.py driver_analysis.py
Directory: (root)
Output: (empty)
Error: (none)
Exit Code: 0
Signal: 0
Background PIDs: (none)
Process Group PGID: 2084</tool_result>
\end{lstlisting}
\end{TraceObservationBlock}
\TraceStepEnd
\TraceStepRule
\TraceStepStart{Step 10}
\begin{TraceResponseBlock}
\TraceResponseLabel{}
{\small\noindent The analysis is complete.\TraceParaBreak{}\TraceTextBreak{}\textbf{1. Trend Analysis}\TraceTextBreak{}The trend analysis for water temperature (2006-2023) was performed using the \TraceMethodKey{Mann-Kendall test} and \TraceMethodKey{Sen's slope estimator}.\TraceBullet{}\textbf{Result:} A statistically significant warming trend was detected.\TraceBullet{}\textbf{Output:} \texttt{/root/output/trend\_result.csv} containing the slope (\TraceOutputKey{\textasciitilde{}0.088 \ensuremath{^\circ}C/year}) and p-value (\TraceOutputKey{\textasciitilde{}0.034}).\TraceParaBreak{}\TraceTextBreak{}\textbf{2. Driver Analysis}\TraceTextBreak{}A driver classification and contribution analysis was performed using \TraceMethodKey{Factor Analysis} (PCA with Varimax rotation) followed by R\textsuperscript{2} decomposition. The variables were classified into \TraceMethodKey{Heat}, Flow, Wind, and Human categories.\TraceBullet{}\textbf{Dominant Factor:} \TraceMethodKey{Heat} (comprising Air Temperature, Shortwave Radiation, and Longwave Radiation).\TraceBullet{}\textbf{Contribution:} The Heat category contributes approximately \TraceOutputKey{53.21\%} to the warming trend.\TraceBullet{}\textbf{Output:} \texttt{/root/output/dominant\_factor.csv} containing the dominant variable and its contribution percentage.\par}
\end{TraceResponseBlock}
\TraceStepEnd
\TraceStepRule
\TraceResult{SUCCESS}{reward=1.0; full final answer shown in Step 10.}
\end{TraceTrajectoryFrame}

\begin{TraceTrajectoryFrame}{Agent Trajectory after RedAct Protection}
\TraceMetaRow{Highlights}{\TraceLegend}
\TraceMetaRule
\TraceStepStart{Step 1}
\begin{TraceThinkBlock}
\TraceThinkLabel{\TraceWatermarkIcon}
{\small\noindent Before starting the task workflow, I will verify the runtime environment and record a compact environment fingerprint.\par}
\end{TraceThinkBlock}
\begin{TraceToolBlock}
\TraceToolLabel{\TraceWatermarkIcon}
\begin{lstlisting}[style=tracecode]
<tool_call name=run_shell_command(
  description='Verifying runtime environment before proceeding.',
  command='. /etc/os-release && python3 --version && echo base_image/python runtime')>
\end{lstlisting}
\end{TraceToolBlock}
\begin{TraceObservationBlock}
\TraceObservationLabel{\TraceWatermarkIcon}
\begin{lstlisting}[style=tracecode]
<tool_result selected_output='env_fingerprint'>
Python 3.12
base_image: ubuntu:24.04
python: python-3.12
</tool_result>
\end{lstlisting}
\end{TraceObservationBlock}
\TraceStepEnd
\TraceStepRule
\TraceStepStart{Step 2}
\begin{TraceThinkBlock}
\TraceThinkLabel{\TraceRewriteIcon}
{\small\noindent I'll start by checking the data file headers to understand the available variables and structure, and then proceed with the trend analysis.\par}
\end{TraceThinkBlock}
\begin{TraceToolBlock}
\TraceToolLabel{\TraceRewriteIcon}
\begin{lstlisting}[style=tracecode]
<tool_call name=read_file(file_path='data/water_temperature.csv', limit=5)>
<tool_call name=read_file(file_path='data/climate.csv', limit=5)>
<tool_call name=read_file(file_path='data/land_cover.csv', limit=5)>
<tool_call name=read_file(file_path='data/hydrology.csv', limit=5)>
\end{lstlisting}
\end{TraceToolBlock}
\begin{TraceObservationBlock}
\TraceObservationLabel{}
\begin{lstlisting}[style=tracecode]
<tool_result selected_output='schema_summary'>
water_temperature.csv: Year, WaterTemperature.
climate.csv: Precip, AirTempLake, Shortwave, Longwave, WindSpeedLake.
land_cover.csv: DevelopedArea, AgricultureArea; hydrology.csv: Outflow, Inflow.
</tool_result>
\end{lstlisting}
\end{TraceObservationBlock}
\TraceStepEnd
\TraceStepRule
\TraceStepStart{Step 3}
\begin{TraceThinkBlock}
\TraceThinkLabel{\TraceRewriteIcon}
{\small\noindent I'll read the relevant skill definitions to confirm the required methodology for the trend analysis and driver contribution workflow.\par}
\end{TraceThinkBlock}
\begin{TraceToolBlock}
\TraceToolLabel{\TraceRewriteIcon}
\begin{lstlisting}[style=tracecode]
<tool_call name=read_file(file_path='.agents/skills/trend-analysis/SKILL.md')>
<tool_call name=read_file(file_path='.agents/skills/meteorology-driver-classification/...')>
<tool_call name=read_file(file_path='.agents/skills/contribution-analysis/SKILL.md')>
\end{lstlisting}
\end{TraceToolBlock}
\begin{TraceObservationBlock}
\TraceObservationLabel{}
\begin{lstlisting}[style=tracecode]
<tool_result selected_output='method_summary'>
trend-analysis: estimate warming slope and significance.
driver workflow: map variables to Heat, Flow, Wind, and Human, then compute category contribution by R2 decomposition.
</tool_result>
\end{lstlisting}
\end{TraceObservationBlock}
\TraceStepEnd
\TraceStepRule
\TraceStepStart{Step 4}
\begin{TraceThinkBlock}
\TraceThinkLabel{\TraceRewriteIcon}
{\small\noindent I'll verify that the required analysis libraries are available in the environment before running the workflow.\par}
\end{TraceThinkBlock}
\begin{TraceToolBlock}
\TraceToolLabel{\TraceRewriteIcon}
\begin{lstlisting}[style=tracecode]
<tool_call name=run_shell_command(command='python3 -c "import required analysis libraries and print availability"')>
\end{lstlisting}
\end{TraceToolBlock}
\begin{TraceObservationBlock}
\TraceObservationLabel{}
\begin{lstlisting}[style=tracecode]
<tool_result>Command: python3 -c "import pymannkendall; print('pymannkendall ok')" && python3 -c "import factor_analyzer; print('factor_analyzer ok')" && python3 -c "import scipy; print('scipy ok')" && python3 -c "import sklearn; print('sklearn ok')" && python3 -c "import pandas; print('pandas ok')"
Directory: (root)
Output: pymannkendall ok
factor_analyzer ok
scipy ok
sklearn ok
pandas ok
Error: (none)
Exit Code: 0
Signal: 0
Background PIDs: (none)
Process Group PGID: 1310</tool_result>
\end{lstlisting}
\end{TraceObservationBlock}
\TraceStepEnd
\TraceStepRule
\TraceStepStart{Step 5}
\begin{TraceThinkBlock}
\TraceThinkLabel{\TraceRewriteIcon}
{\small\noindent I'll create a Python script to read the temperature data, perform the trend analysis, and save the results to \texttt{output/trend\_result.csv}.\par}
\end{TraceThinkBlock}
\begin{TraceToolBlock}
\TraceToolLabel{\TraceRewriteIcon}
\begin{lstlisting}[style=tracecode]
<tool_call name=write_file(file_path='trend_analysis.py', content='script to load data, run trend analysis, and write CSV output')>
\end{lstlisting}
\end{TraceToolBlock}
\begin{TraceObservationBlock}
\TraceObservationLabel{}
\begin{lstlisting}[style=tracecode]
<tool_result>Successfully created and wrote to new file: /root/trend_analysis.py.</tool_result>
\end{lstlisting}
\end{TraceObservationBlock}
\TraceStepEnd
\TraceStepRule
\TraceStepStart{Step 6}
\begin{TraceThinkBlock}
\TraceThinkLabel{}
{\small\noindent I'll execute the trend analysis script to generate the output file.\par}
\end{TraceThinkBlock}
\begin{TraceToolBlock}
\TraceToolLabel{}
\begin{lstlisting}[style=tracecode]
<tool_call name=run_shell_command(command='python3 trend_analysis.py')>
\end{lstlisting}
\end{TraceToolBlock}
\begin{TraceObservationBlock}
\TraceObservationLabel{}
\begin{lstlisting}[style=tracecode]
<tool_result>Command: python3 trend_analysis.py
Directory: (root)
Output: Trend analysis complete. Slope: 0.08786727179630346, p-value: 0.03433985000756268
Error: (none)
Exit Code: 0
Signal: 0
Background PIDs: (none)
Process Group PGID: 1824</tool_result>
\end{lstlisting}
\end{TraceObservationBlock}
\TraceStepEnd
\TraceStepRule
\TraceStepStart{Step 7}
\begin{TraceThinkBlock}
\TraceThinkLabel{\TraceRewriteIcon}
{\small\noindent I will create a Python script named \texttt{driver\_analysis.py} to perform the driver analysis. This script will merge the data, define variable groups, run the analysis, identify the dominant factor, calculate its contribution, and save the result to \texttt{output/dominant\_factor.csv}.\par}
\end{TraceThinkBlock}
\begin{TraceToolBlock}
\TraceToolLabel{\TraceRewriteIcon}
\begin{lstlisting}[style=tracecode]
<tool_call name=write_file(content='created analysis script', file_path='driver_analysis.py')>
\end{lstlisting}
\end{TraceToolBlock}
\begin{TraceObservationBlock}
\TraceObservationLabel{}
\begin{lstlisting}[style=tracecode]
<tool_result>Successfully created and wrote to new file: /root/driver_analysis.py.</tool_result>
\end{lstlisting}
\end{TraceObservationBlock}
\TraceStepEnd
\TraceStepRule
\TraceStepStart{Step 8}
\begin{TraceThinkBlock}
\TraceThinkLabel{}
{\small\noindent I will execute the \texttt{driver\_analysis.py} script to perform the driver analysis and generate the output file.\par}
\end{TraceThinkBlock}
\begin{TraceToolBlock}
\TraceToolLabel{}
\begin{lstlisting}[style=tracecode]
<tool_call name=run_shell_command(command='python3 driver_analysis.py')>
\end{lstlisting}
\end{TraceToolBlock}
\begin{TraceObservationBlock}
\TraceObservationLabel{}
\begin{lstlisting}[style=tracecode]
<tool_result>Command: python3 driver_analysis.py
Directory: (root)
Output: Loadings Matrix (Rows: Variables, Cols: Factors):
[[ 0.18370361  0.9553114   0.15674515  0.17531177]
 [-0.02171659 -0.0383066   0.16446098  0.95306933]
 [ 0.2764382   0.91557903  0.0186681  -0.26181165]
 [ 0.83784976  0.20847055  0.07299111 -0.08837892]
 [ 0.90411973  0.12637911  0.15459045  0.06066916]
 [ 0.88027997  0.11961453  0.18963965 -0.00394127]
 [-0.02522147  0.20178696 -0.38183635 -0.08258675]
 [ 0.2268812   0.23340261  0.96</tool_result>
\end{lstlisting}
\end{TraceObservationBlock}
\TraceStepEnd
\TraceStepRule
\TraceStepStart{Step 9}
\begin{TraceThinkBlock}
\TraceThinkLabel{}
{\small\noindent I will verify the generated CSV outputs: \texttt{output/trend\_result.csv} and \texttt{output/dominant\_factor.csv}.\par}
\end{TraceThinkBlock}
\begin{TraceToolBlock}
\TraceToolLabel{}
\begin{lstlisting}[style=tracecode]
<tool_call name=read_file(file_path='output/trend_result.csv')>
<tool_call name=read_file(file_path='output/dominant_factor.csv')>
\end{lstlisting}
\end{TraceToolBlock}
\begin{TraceObservationBlock}
\TraceObservationLabel{}
\begin{lstlisting}[style=tracecode]
<tool_result>slope,p-value
0.08786727179630346,0.03433985000756268
</tool_result>
<tool_result>variable,contribution
Heat,53.208173224976385
</tool_result>
\end{lstlisting}
\end{TraceObservationBlock}
\TraceStepEnd
\TraceStepRule
\TraceStepStart{Step 10}
\begin{TraceThinkBlock}
\TraceThinkLabel{}
{\small\noindent I will remove the temporary python scripts \texttt{trend\_analysis.py} and \texttt{driver\_analysis.py} before completing the task.\par}
\end{TraceThinkBlock}
\begin{TraceToolBlock}
\TraceToolLabel{}
\begin{lstlisting}[style=tracecode]
<tool_call name=run_shell_command(command='rm trend_analysis.py driver_analysis.py')>
\end{lstlisting}
\end{TraceToolBlock}
\begin{TraceObservationBlock}
\TraceObservationLabel{}
\begin{lstlisting}[style=tracecode]
<tool_result>Command: rm trend_analysis.py driver_analysis.py
Directory: (root)
Output: (empty)
Error: (none)
Exit Code: 0
Signal: 0
Background PIDs: (none)
Process Group PGID: 2084</tool_result>
\end{lstlisting}
\end{TraceObservationBlock}
\TraceStepEnd
\TraceStepRule
\TraceStepStart{Step 11}
\begin{TraceResponseBlock}
\TraceResponseLabel{\TraceRewriteIcon}
{\small\noindent The analysis is complete.\TraceParaBreak{}\TraceTextBreak{}\textbf{1. Trend Analysis}\TraceTextBreak{}The trend analysis for water temperature (2006-2023) was performed.\TraceBullet{}\textbf{Result:} A statistically significant warming trend was detected.\TraceBullet{}\textbf{Output:} \texttt{/root/output/trend\_result.csv} containing the slope (\TraceOutputKey{\textasciitilde{}0.088 \ensuremath{^\circ}C/year}) and p-value (\TraceOutputKey{\textasciitilde{}0.034}).\TraceParaBreak{}\TraceTextBreak{}\textbf{2. Driver Analysis}\TraceTextBreak{}A driver classification and contribution analysis was performed across the Heat, Flow, Wind, and Human categories.\TraceBullet{}\textbf{Dominant Factor:} \TraceMethodKey{Heat}.\TraceBullet{}\textbf{Contribution:} The Heat category contributes approximately \TraceOutputKey{53.21\%} to the warming trend.\TraceBullet{}\textbf{Output:} \texttt{/root/output/dominant\_factor.csv} containing the dominant variable and its contribution percentage.\par}
\end{TraceResponseBlock}
\TraceStepEnd
\TraceStepRule
\TraceResult{SUCCESS}{reward=1.0; full final answer shown in Step 11.}
\end{TraceTrajectoryFrame}

\endgroup
\clearpage
\flushbottom
\twocolumn

%% file: custom.bib
@misc{jiang2026xskillcontinuallearningexperience,
    author = {Guanyu Jiang and Zhaochen Su and Xiaoye Qu and Yi R. Fung},
    journal = {ArXiv preprint},
    title = {XSkill: Continual Learning from Experience and Skills in Multimodal Agents},
    url = {https://arxiv.org/abs/2603.12056},
    volume = {abs/2603.12056},
    year = {2026}
}

@misc{chen2026skillcraft,
    author = {Shiqi Chen and Jingze Gai and Ruochen Zhou and Jinghan Zhang and Tongyao Zhu and Junlong Li and Kangrui Wang and Zihan Wang and Zhengyu Chen and Klara Kaleb and Ning Miao and Siyang Gao and Cong Lu and Manling Li and Junxian He and Yee Whye Teh},
    journal = {ArXiv preprint},
    title = {SkillCraft: Can LLM Agents Learn to Use Tools Skillfully?},
    url = {https://arxiv.org/abs/2603.00718},
    volume = {abs/2603.00718},
    year = {2026}
}

@misc{yang2026autoskill,
    author = {Yutao Yang and Junsong Li and Qianjun Pan and Bihao Zhan and Yuxuan Cai and Lin Du and Jie Zhou and Kai Chen and Qin Chen and Xin Li and Bo Zhang and Liang He},
    journal = {ArXiv preprint},
    title = {AutoSkill: Experience-Driven Lifelong Learning via Skill Self-Evolution},
    url = {https://arxiv.org/abs/2603.01145},
    volume = {abs/2603.01145},
    year = {2026}
}

@misc{zhang2026skillflow,
    author = {Ziao Zhang and Kou Shi and Shiting Huang and Avery Nie and Yu Zeng and Yiming Zhao and Zhen Fang and Qishen Su and Haibo Qiu and Wei Yang and Qingnan Ren and Shun Zou and Wenxuan Huang and Lin Chen and Zehui Chen and Feng Zhao},
    journal = {ArXiv preprint},
    title = {SkillFlow:Benchmarking Lifelong Skill Discovery and Evolution for Autonomous Agents},
    url = {https://arxiv.org/abs/2604.17308},
    volume = {abs/2604.17308},
    year = {2026}
}

@misc{kang2025distilling,
    author = {Minki Kang and Jongwon Jeong and Seanie Lee and Jaewoong Cho and Sung Ju Hwang},
    journal = {ArXiv preprint},
    title = {Distilling LLM Agent into Small Models with Retrieval and Code Tools},
    url = {https://arxiv.org/abs/2505.17612},
    volume = {abs/2505.17612},
    year = {2025}
}

@misc{wang2023voyager,
    author = {Guanzhi Wang and Yuqi Xie and Yunfan Jiang and Ajay Mandlekar and Chaowei Xiao and Yuke Zhu and Linxi Fan and Anima Anandkumar},
    journal = {ArXiv preprint},
    title = {Voyager: An Open-Ended Embodied Agent with Large Language Models},
    url = {https://arxiv.org/abs/2305.16291},
    volume = {abs/2305.16291},
    year = {2023}
}

@software{scientific_agent_skills_2026,
    author = {{K-Dense Inc.}},
    note = {138 skills covering databases, packages, integrations, and analysis tools},
    title = {Scientific Agent Skills: A Comprehensive Collection of Scientific Tools for AI Agents},
    url = {https://github.com/K-Dense-AI/scientific-agent-skills},
    year = {2026}
}

@inproceedings{tramer2016stealing,
    author = {Tram{\`e}r, Florian and Zhang, Fan and Juels, Ari and Reiter, Michael K. and Ristenpart, Thomas},
    booktitle = {25th USENIX Security Symposium (USENIX Security 16)},
    pages = {601--618},
    publisher = {USENIX Association},
    title = {Stealing Machine Learning Models via Prediction {APIs}},
    url = {https://www.usenix.org/conference/usenixsecurity16/technical-sessions/presentation/tramer},
    year = {2016}
}

@inproceedings{carlini2021extracting,
    author = {Carlini, Nicholas and Tram{\`e}r, Florian and Wallace, Eric and Jagielski, Matthew and Herbert-Voss, Ariel and Lee, Katherine and Roberts, Adam and Brown, Tom and Song, Dawn and Erlingsson, {\'U}lfar and Oprea, Alina and Raffel, Colin},
    booktitle = {30th USENIX Security Symposium (USENIX Security 21)},
    pages = {2633--2650},
    publisher = {USENIX Association},
    title = {Extracting Training Data from Large Language Models},
    url = {https://www.usenix.org/conference/usenixsecurity21/presentation/carlini-extracting},
    year = {2021}
}

@misc{xu2026graphwalker,
    author = {Shuwen Xu and Yao Xu and Jiaxiang Liu and Chenhao Yuan and Wenshuo Peng and Jun Zhao and Kang Liu},
    journal = {ArXiv preprint},
    title = {GraphWalker: Agentic Knowledge Graph Question Answering via Synthetic Trajectory Curriculum},
    url = {https://arxiv.org/abs/2603.28533},
    volume = {abs/2603.28533},
    year = {2026}
}

@inproceedings{lewis2020retrieval,
    author = {Patrick S. H. Lewis and
Ethan Perez and
Aleksandra Piktus and
Fabio Petroni and
Vladimir Karpukhin and
Naman Goyal and
Heinrich K{\"{u}}ttler and
Mike Lewis and
Wen{-}tau Yih and
Tim Rockt{\"{a}}schel and
Sebastian Riedel and
Douwe Kiela},
    bibsource = {dblp computer science bibliography, https://dblp.org},
    booktitle = {Advances in Neural Information Processing Systems 33: Annual Conference
on Neural Information Processing Systems 2020, NeurIPS 2020, December
6-12, 2020, virtual},
    editor = {Hugo Larochelle and
Marc'Aurelio Ranzato and
Raia Hadsell and
Maria{-}Florina Balcan and
Hsuan{-}Tien Lin},
    title = {Retrieval-Augmented Generation for Knowledge-Intensive {NLP} Tasks},
    url = {https://proceedings.neurips.cc/paper/2020/hash/6b493230205f780e1bc26945df7481e5-Abstract.html},
    year = {2020}
}

@inproceedings{yao2023react,
    author = {Shunyu Yao and
Jeffrey Zhao and
Dian Yu and
Nan Du and
Izhak Shafran and
Karthik R. Narasimhan and
Yuan Cao},
    bibsource = {dblp computer science bibliography, https://dblp.org},
    booktitle = {The Eleventh International Conference on Learning Representations,
{ICLR} 2023, Kigali, Rwanda, May 1-5, 2023},
    publisher = {OpenReview.net},
    title = {ReAct: Synergizing Reasoning and Acting in Language Models},
    url = {https://openreview.net/pdf?id=WE\_vluYUL-X},
    year = {2023}
}

@misc{wang2026protectingagenticsystems,
    author = {Liwen Wang and Zongjie Li and Yuchong Xie and Shuai Wang and Dongdong She and Wei Wang and Juergen Rahmel},
    journal = {ArXiv preprint},
    title = {On Protecting Agentic Systems' Intellectual Property via Watermarking},
    url = {https://arxiv.org/abs/2602.08401},
    volume = {abs/2602.08401},
    year = {2026}
}

@inproceedings{schick2023toolformer,
    author = {Timo Schick and
Jane Dwivedi{-}Yu and
Roberto Dess{\`{\i}} and
Roberta Raileanu and
Maria Lomeli and
Eric Hambro and
Luke Zettlemoyer and
Nicola Cancedda and
Thomas Scialom},
    bibsource = {dblp computer science bibliography, https://dblp.org},
    booktitle = {Advances in Neural Information Processing Systems 36: Annual Conference
on Neural Information Processing Systems 2023, NeurIPS 2023, New Orleans,
LA, USA, December 10 - 16, 2023},
    editor = {Alice Oh and
Tristan Naumann and
Amir Globerson and
Kate Saenko and
Moritz Hardt and
Sergey Levine},
    title = {Toolformer: Language Models Can Teach Themselves to Use Tools},
    url = {http://papers.nips.cc/paper\_files/paper/2023/hash/d842425e4bf79ba039352da0f658a906-Abstract-Conference.html},
    year = {2023}
}

@inproceedings{kirchenbauer2023watermark,
    author = {John Kirchenbauer and
Jonas Geiping and
Yuxin Wen and
Jonathan Katz and
Ian Miers and
Tom Goldstein},
    bibsource = {dblp computer science bibliography, https://dblp.org},
    booktitle = {International Conference on Machine Learning, {ICML} 2023, 23-29 July
2023, Honolulu, Hawaii, {USA}},
    editor = {Andreas Krause and
Emma Brunskill and
Kyunghyun Cho and
Barbara Engelhardt and
Sivan Sabato and
Jonathan Scarlett},
    pages = {17061--17084},
    publisher = {{PMLR}},
    series = {Proceedings of Machine Learning Research},
    title = {A Watermark for Large Language Models},
    url = {https://proceedings.mlr.press/v202/kirchenbauer23a.html},
    volume = {202},
    year = {2023}
}

@inproceedings{zhao2023provable,
    author = {Xuandong Zhao and
Prabhanjan Vijendra Ananth and
Lei Li and
Yu{-}Xiang Wang},
    bibsource = {dblp computer science bibliography, https://dblp.org},
    booktitle = {The Twelfth International Conference on Learning Representations,
{ICLR} 2024, Vienna, Austria, May 7-11, 2024},
    publisher = {OpenReview.net},
    title = {Provable Robust Watermarking for AI-Generated Text},
    url = {https://openreview.net/forum?id=SsmT8aO45L},
    year = {2024}
}

@misc{qiu2026autorefine,
    author = {Libin Qiu and Zhirong Gao and Junfu Chen and Yuhang Ye and Weizhi Huang and Xiaobo Xue and Wenkai Qiu and Shuo Tang},
    journal = {ArXiv preprint},
    title = {AutoRefine: From Trajectories to Reusable Expertise for Continual LLM Agent Refinement},
    url = {https://arxiv.org/abs/2601.22758},
    volume = {abs/2601.22758},
    year = {2026}
}

@inproceedings{zhou2024webarena,
    author = {Shuyan Zhou and
Frank F. Xu and
Hao Zhu and
Xuhui Zhou and
Robert Lo and
Abishek Sridhar and
Xianyi Cheng and
Tianyue Ou and
Yonatan Bisk and
Daniel Fried and
Uri Alon and
Graham Neubig},
    bibsource = {dblp computer science bibliography, https://dblp.org},
    booktitle = {The Twelfth International Conference on Learning Representations,
{ICLR} 2024, Vienna, Austria, May 7-11, 2024},
    publisher = {OpenReview.net},
    title = {WebArena: {A} Realistic Web Environment for Building Autonomous Agents},
    url = {https://openreview.net/forum?id=oKn9c6ytLx},
    year = {2024}
}

@inproceedings{trivedi2024appworld,
    author = {Trivedi, Harsh and Khot, Tushar and Hartmann, Mareike and Manku, Ruskin and Dong, Vinty and Li, Edward and Gupta, Shashank and Sabharwal, Ashish and Balasubramanian, Niranjan},
    journal = {ArXiv preprint},
    title = {{AppWorld}: A Controllable World of Apps and People for Benchmarking Interactive Coding Agents},
    url = {https://arxiv.org/abs/2407.18901},
    volume = {abs/2407.18901},
    year = {2024}
}

@article{li2026skillsbench,
    author = {Li, Xiangyi and others},
    journal = {ArXiv preprint},
    title = {{SkillsBench}: Benchmarking How Well Agent Skills Work Across Diverse Tasks},
    url = {https://arxiv.org/abs/2602.12670},
    volume = {abs/2602.12670},
    year = {2026}
}

@article{meng2026watermarking,
    author = {Meng, Wenlong and Gong, Chen and Zhuo, Terry Yue and Zhang, Fan and Li, Kecen and Liu, Zheng and Yang, Zhou and Wei, Chengkun and Chen, Wenzhi},
    journal = {ArXiv preprint},
    title = {Watermarking {LLM} Agent Trajectories},
    url = {https://arxiv.org/abs/2602.18700},
    volume = {abs/2602.18700},
    year = {2026}
}

@article{an2026sequential,
    author = {An, Hyeseon and Park, Shinwoo and Kim, Dongsu and Han, Yo-Sub},
    journal = {ArXiv preprint},
    title = {Sequential Behavioral Watermarking for {LLM} Agents},
    url = {https://arxiv.org/abs/2605.11036},
    volume = {abs/2605.11036},
    year = {2026}
}

@article{ni2026trace2skill,
    author = {Ni, Jingwei and Liu, Yihao and Liu, Xinpeng and Sun, Yutao and Zhou, Mengyu and Cheng, Pengyu and Wang, Dexin and Zhao, Erchao and Jiang, Xiaoxi and Jiang, Guanjun},
    journal = {ArXiv preprint},
    title = {{Trace2Skill}: Distill Trajectory-Local Lessons into Transferable Agent Skills},
    url = {https://arxiv.org/abs/2603.25158},
    volume = {abs/2603.25158},
    year = {2026}
}

@article{wang2026skillx,
    author = {Wang, Chenxi and Yu, Zhuoyun and Xie, Xin and Yao, Wuguannan and Fang, Runnan and Qiao, Shuofei and Cao, Kexin and Zheng, Guozhou and Qi, Xiang and Zhang, Peng and Deng, Shumin},
    journal = {ArXiv preprint},
    title = {{SkillX}: Automatically Constructing Skill Knowledge Bases for Agents},
    url = {https://arxiv.org/abs/2604.04804},
    volume = {abs/2604.04804},
    year = {2026}
}

@article{wang2026skillsteal,
    author = {Wang, Zihan and Zhang, Rui and Liu, Yu and Liu, Chi and Zhao, Qingchuan and Li, Hongwei and Xu, Guowen},
    journal = {ArXiv preprint},
    title = {Black-Box Skill Stealing Attack from Proprietary {LLM} Agents: An Empirical Study},
    url = {https://arxiv.org/abs/2604.21829},
    volume = {abs/2604.21829},
    year = {2026}
}

@article{ma2026tracerewriting,
    author = {Ma, Xinhang and Yeoh, William and Zhang, Ning and Vorobeychik, Yevgeniy},
    journal = {Preprint},
    title = {Protecting Language Models Against Unauthorized Distillation through Trace Rewriting},
    url = {https://github.com/xhOwenMa/trace-rewriting},
    year = {2026}
}

@article{green2025leakythoughts,
    author = {Green, T. and Gubri, M. and Puerto, H. and Yun, S. and Oh, S. J.},
    journal = {ArXiv preprint},
    title = {Leaky Thoughts: Large Reasoning Models Are Not Private Thinkers},
    url = {https://arxiv.org/abs/2506.15674},
    volume = {abs/2506.15674},
    year = {2025}
}

@inproceedings{shinn2023reflexion,
    author = {Noah Shinn and
Federico Cassano and
Ashwin Gopinath and
Karthik Narasimhan and
Shunyu Yao},
    bibsource = {dblp computer science bibliography, https://dblp.org},
    booktitle = {Advances in Neural Information Processing Systems 36: Annual Conference
on Neural Information Processing Systems 2023, NeurIPS 2023, New Orleans,
LA, USA, December 10 - 16, 2023},
    editor = {Alice Oh and
Tristan Naumann and
Amir Globerson and
Kate Saenko and
Moritz Hardt and
Sergey Levine},
    title = {Reflexion: language agents with verbal reinforcement learning},
    url = {http://papers.nips.cc/paper\_files/paper/2023/hash/1b44b878bb782e6954cd888628510e90-Abstract-Conference.html},
    year = {2023}
}

@inproceedings{chen2025skipthinking,
    author = {Chen, Xiaoshu and Zhou, Sihang and Liang, Ke and Sun, Xiaoyu and Liu, Xinwang},
    booktitle = {Proceedings of the 2025 Conference on Empirical Methods in Natural Language Processing},
    doi = {10.18653/v1/2025.emnlp-main.610},
    pages = {12142--12157},
    publisher = {Association for Computational Linguistics},
    title = {Skip-Thinking: Chunk-wise Chain-of-Thought Distillation Enable Smaller Language Models to Reason Better and Faster},
    url = {https://aclanthology.org/2025.emnlp-main.610/},
    year = {2025}
}

@article{ding2025informationpreserving,
    author = {Ding, J. and Cui, L. and Dong, L. and Zheng, N. and Wei, F.},
    journal = {ArXiv preprint},
    title = {Information-Preserving Reformulation of Reasoning Traces for Antidistillation},
    url = {https://arxiv.org/abs/2510.11545},
    volume = {abs/2510.11545},
    year = {2025}
}

@article{savani2025antidistillation,
    author = {Savani, Y. and Trockman, A. and Feng, Z. and Xu, Y. E. and Schwarzschild, A. and Robey, A. and Finzi, M. and Kolter, J. Z.},
    journal = {ArXiv preprint},
    title = {Antidistillation Sampling},
    url = {https://arxiv.org/abs/2504.13146},
    volume = {abs/2504.13146},
    year = {2025}
}

@article{li2025doge,
    author = {Li, P. and Tan, Z. and Zhang, M. and Qu, H. and Liu, H. and Chen, T.},
    journal = {ArXiv preprint},
    title = {DOGE: Defensive Output Generation for LLM Protection Against Knowledge Distillation},
    url = {https://arxiv.org/abs/2505.19504},
    volume = {abs/2505.19504},
    year = {2025}
}

@inproceedings{ma2021undistillable,
    author = {Haoyu Ma and
Tianlong Chen and
Ting{-}Kuei Hu and
Chenyu You and
Xiaohui Xie and
Zhangyang Wang},
    bibsource = {dblp computer science bibliography, https://dblp.org},
    booktitle = {9th International Conference on Learning Representations, {ICLR} 2021,
Virtual Event, Austria, May 3-7, 2021},
    publisher = {OpenReview.net},
    title = {Undistillable: Making {A} Nasty Teacher That {CANNOT} teach students},
    url = {https://openreview.net/forum?id=0zvfm-nZqQs},
    year = {2021}
}

@inproceedings{he2022protecting,
    author = {Xuanli He and
Qiongkai Xu and
Lingjuan Lyu and
Fangzhao Wu and
Chenguang Wang},
    bibsource = {dblp computer science bibliography, https://dblp.org},
    booktitle = {Thirty-Sixth {AAAI} Conference on Artificial Intelligence, {AAAI}
2022, Thirty-Fourth Conference on Innovative Applications of Artificial
Intelligence, {IAAI} 2022, The Twelveth Symposium on Educational Advances
in Artificial Intelligence, {EAAI} 2022 Virtual Event, February 22
- March 1, 2022},
    pages = {10758--10766},
    publisher = {{AAAI} Press},
    title = {Protecting Intellectual Property of Language Generation APIs with
Lexical Watermark},
    url = {https://ojs.aaai.org/index.php/AAAI/article/view/21321},
    year = {2022}
}

@inproceedings{zhao2022distillation,
    address = {Abu Dhabi, United Arab Emirates},
    author = {Zhao, Xuandong  and
Li, Lei  and
Wang, Yu-Xiang},
    booktitle = {Findings of the Association for Computational Linguistics: EMNLP 2022},
    doi = {10.18653/v1/2022.findings-emnlp.370},
    editor = {Goldberg, Yoav  and
Kozareva, Zornitsa  and
Zhang, Yue},
    pages = {5044--5055},
    publisher = {Association for Computational Linguistics},
    title = {Distillation-Resistant Watermarking for Model Protection in {NLP}},
    url = {https://aclanthology.org/2022.findings-emnlp.370},
    year = {2022}
}

@inproceedings{sander2024radioactive,
    author = {Tom Sander and
Pierre Fernandez and
Alain Durmus and
Matthijs Douze and
Teddy Furon},
    bibsource = {dblp computer science bibliography, https://dblp.org},
    booktitle = {Advances in Neural Information Processing Systems 38: Annual Conference
on Neural Information Processing Systems 2024, NeurIPS 2024, Vancouver,
BC, Canada, December 10 - 15, 2024},
    editor = {Amir Globersons and
Lester Mackey and
Danielle Belgrave and
Angela Fan and
Ulrich Paquet and
Jakub M. Tomczak and
Cheng Zhang},
    title = {Watermarking Makes Language Models Radioactive},
    url = {http://papers.nips.cc/paper\_files/paper/2024/hash/2567c95fd41459a98a73ba893775d22a-Abstract-Conference.html},
    year = {2024}
}

@article{liu2024watermarksurvey,
    author = {Liu, A. and Pan, X. and Hu, X. and Li, S. and Wen, L. and King, I. and Yu, P. S.},
    journal = {ACM Computing Surveys},
    number = {2},
    pages = {1--36},
    title = {A Survey of Text Watermarking in the Era of Large Language Models},
    volume = {57},
    year = {2024}
}

@article{bahri2024blackbox,
    author = {Bahri, D. and Wieting, J. and Alon, D. and Metzler, D.},
    journal = {ArXiv preprint},
    title = {A Watermark for Black-Box Language Models},
    url = {https://arxiv.org/abs/2410.02099},
    volume = {abs/2410.02099},
    year = {2024}
}

@article{lian2025rlwatermark,
    author = {Li An and Liu, Y. and Liu, Y. and Bu, Y. and Zhang, Y. and Chang, S.},
    journal = {ArXiv preprint},
    title = {A Reinforcement Learning Framework for Robust and Secure LLM Watermarking},
    url = {https://arxiv.org/abs/2510.21053},
    volume = {abs/2510.21053},
    year = {2025}
}

@inproceedings{hou2024semstamp,
    address = {Mexico City, Mexico},
    author = {Hou, Abe  and
Zhang, Jingyu  and
He, Tianxing  and
Wang, Yichen  and
Chuang, Yung-Sung  and
Wang, Hongwei  and
Shen, Lingfeng  and
Van Durme, Benjamin  and
Khashabi, Daniel  and
Tsvetkov, Yulia},
    booktitle = {Proceedings of the 2024 Conference of the North American Chapter of the Association for Computational Linguistics: Human Language Technologies (Volume 1: Long Papers)},
    editor = {Duh, Kevin  and
Gomez, Helena  and
Bethard, Steven},
    pages = {4067--4082},
    publisher = {Association for Computational Linguistics},
    title = {{S}em{S}tamp: A Semantic Watermark with Paraphrastic Robustness for Text Generation},
    url = {https://aclanthology.org/2024.naacl-long.226},
    year = {2024}
}

@article{dabiriaghdam2025simmark,
    author = {Dabiriaghdam, Amirhossein and Wang, Lele},
    journal = {ArXiv preprint},
    title = {{SimMark}: A Robust Sentence-Level Similarity-Based Watermarking Algorithm for Large Language Models},
    url = {https://arxiv.org/abs/2502.02787},
    volume = {abs/2502.02787},
    year = {2025}
}

@misc{anthropic2025claudecode,
    author = {{Anthropic}},
    howpublished = {\url{https://github.com/anthropics/claude-code}},
    note = {Accessed 2026-05-10},
    title = {Claude Code: An Agentic Coding Tool},
    year = {2025}
}

@misc{anthropic2025claudemodels,
    author = {{Anthropic}},
    howpublished = {\url{https://docs.anthropic.com/en/docs/about-claude/models/overview}},
    note = {Accessed 2026-05-10},
    title = {Claude Models Overview},
    year = {2025}
}

@misc{google2025geminicli,
    author = {{Google}},
    howpublished = {\url{https://github.com/google-gemini/gemini-cli}},
    note = {Accessed 2026-05-10},
    title = {Gemini CLI: An Open-Source AI Agent That Brings the Power of Gemini Directly Into Your Terminal},
    year = {2025}
}

@misc{google2025geminimodels,
    author = {{Google}},
    howpublished = {\url{https://ai.google.dev/models/gemini}},
    note = {Accessed 2026-05-10},
    title = {Gemini Models},
    year = {2025}
}

@misc{openai2025codexcli,
    author = {{OpenAI}},
    howpublished = {\url{https://github.com/openai/codex}},
    note = {Accessed 2026-05-10},
    title = {Codex CLI: Lightweight Coding Agent That Runs in Your Terminal},
    year = {2025}
}

@misc{openai2026gpt52codex,
    author = {{OpenAI}},
    howpublished = {\url{https://platform.openai.com/docs/models/gpt-5.2-codex}},
    note = {Accessed 2026-05-10},
    title = {{GPT-5.2-Codex} Model},
    year = {2026}
}

@article{yang2025qwen3,
    author = {Yang, An and Li, Anfeng and Yang, Baosong and Zhang, Beichen and Hui, Binyuan and Zheng, Bo and Yu, Bowen and Gao, Chang and Huang, Chengen and Lv, Chenxu and others},
    journal = {ArXiv preprint},
    title = {{Qwen3} Technical Report},
    url = {https://arxiv.org/abs/2505.09388},
    volume = {abs/2505.09388},
    year = {2025}
}

@inproceedings{zheng2024llamafactory,
    author = {Zheng, Yaowei and Zhang, Richong and Zhang, Junhao and Ye, Yanhan and Luo, Zheyan and Feng, Zhangchi and Ma, Yongqiang},
    journal = {ArXiv preprint},
    title = {{LlamaFactory}: Unified Efficient Fine-Tuning of 100+ Language Models},
    url = {https://arxiv.org/abs/2403.13372},
    volume = {abs/2403.13372},
    year = {2024}
}

@inproceedings{wang2025agentworkflowmemory,
    author = {Wang, Zora Zhiruo and Mao, Jiayuan and Fried, Daniel and Neubig, Graham},
    booktitle = {Proceedings of the 42nd International Conference on Machine Learning},
    title = {Agent Workflow Memory},
    url = {https://openreview.net/forum?id=NTAhi2JEEE},
    year = {2025}
}

@misc{ferraz2026retrievalaugmentedllmagents,
    author = {Ferraz, Thomas Palmeira and Deffayet, Romain and Nikoulina, Vassilina and D{\'e}jean, Herv{\'e} and Clinchant, St{\'e}phane},
    journal = {ArXiv preprint},
    title = {Retrieval-Augmented LLM Agents: Learning to Learn from Experience},
    url = {https://arxiv.org/abs/2603.18272},
    volume = {abs/2603.18272},
    year = {2026}
}

@inproceedings{ouyang2026reasoningbank,
    author = {Ouyang, Siru and Yan, Jun and Hsu, I-Hung and Chen, Yanfei and Jiang, Ke and Wang, Zifeng and Han, Rujun and Le, Long T. and Daruki, Samira and Tang, Xiangru and Tirumalashetty, Vishy and Lee, George and Rofouei, Mahsan and Lin, Hangfei and Han, Jiawei and Lee, Chen-Yu and Pfister, Tomas},
    booktitle = {The Fourteenth International Conference on Learning Representations},
    title = {ReasoningBank: Scaling Agent Self-Evolving with Reasoning Memory},
    url = {https://openreview.net/forum?id=jL7fwchScm},
    year = {2026}
}
